\documentclass[10pt,twocolumn,journal]{IEEEtran}
\usepackage[T1]{fontenc}
\usepackage[latin9]{inputenc}
\usepackage{color}
\usepackage{array}
\usepackage{float}
\usepackage{multirow}
\usepackage{amsmath}
\usepackage{amsthm}
\usepackage{amssymb}
\usepackage{stmaryrd}
\usepackage{stackrel}
\usepackage{graphicx}
\usepackage{setspace}
\usepackage[unicode=true,
 bookmarks=true,bookmarksnumbered=true,bookmarksopen=true,bookmarksopenlevel=1,
 breaklinks=false,pdfborder={0 0 0},pdfborderstyle={},backref=false,colorlinks=false]
 {hyperref}
\hypersetup{pdftitle={Your Title},
 pdfauthor={Your Name},
 pdfpagelayout=OneColumn, pdfnewwindow=true, pdfstartview=XYZ, plainpages=false}

\makeatletter

\providecommand{\tabularnewline}{\\}
\floatstyle{ruled}
\newfloat{algorithm}{tbp}{loa}
\providecommand{\algorithmname}{Algorithm}
\floatname{algorithm}{\protect\algorithmname}

\theoremstyle{plain}
\newtheorem{thm}{\protect\theoremname}
\theoremstyle{plain}
\newtheorem{lem}[thm]{\protect\lemmaname}

\usepackage[caption=false,font=footnotesize]{subfig}
\usepackage{algorithmic}
\usepackage{cite}

\@ifundefined{showcaptionsetup}{}{%
 \PassOptionsToPackage{caption=false}{subfig}}
\usepackage{subfig}
\makeatother

\providecommand{\lemmaname}{Lemma}
\providecommand{\theoremname}{Theorem}

\begin{document}
\title{Successive Linear Approximation VBI for Joint Sparse Signal Recovery
and Dynamic Grid Parameters Estimation}
\author{Wenkang~Xu,~An Liu, \IEEEmembership{Senior Member,~IEEE,} Bingpeng
Zhou, and Min-jian Zhao{\normalsize{}}\thanks{Wenkang Xu, An Liu, and Min-jian Zhao are with the College of Information
Science and Electronic Engineering, Zhejiang University, Hangzhou
310027, China (email: anliu@zju.edu.cn).

Bingpeng Zhou is with the School of Electronics and Communication
Engineering, Shenzhen Campus of Sun Yat-sen University, Shenzhen 518000,
China (email:zhoubp3@mail.sysu.edu.cn).}}
\maketitle
\begin{abstract}
For many practical applications in wireless communications, we need
to recover a structured sparse signal from a linear observation model
with dynamic grid parameters in the sensing matrix. Conventional expectation
maximization (EM)-based compressed sensing (CS) methods, such as turbo
compressed sensing (Turbo-CS) and turbo variational Bayesian inference
(Turbo-VBI), have double-loop iterations, where the inner loop (E-step)
obtains a Bayesian estimation of sparse signals and the outer loop
(M-step) obtains a point estimation of dynamic grid parameters. This
leads to a slow convergence rate. Furthermore, each iteration of the
E-step involves a complicated matrix inverse in general. To overcome
these drawbacks, we first propose a successive linear approximation
VBI (SLA-VBI) algorithm that can provide Bayesian estimation of both
sparse signals and dynamic grid parameters. Besides, we simplify the
matrix inverse operation based on the majorization-minimization (MM)
algorithmic framework. In addition, we extend our proposed algorithm
from an independent sparse prior to more complicated structured sparse
priors, which can exploit structured sparsity in specific applications
to further enhance the performance. Finally, we apply our proposed
algorithm to solve two practical application problems in wireless
communications and verify that the proposed algorithm can achieve
faster convergence, lower complexity, and better performance compared
to the state-of-the-art EM-based methods.
\end{abstract}


\section{Introduction}

Compressed sensing (CS) has been widely used in many applications,
such as channel estimation \textcolor{blue}{\cite{Berger_CE,Paredes_CE,ChengLei_Array_manifold}},
data detection \cite{Cirik_DD,Prasad_DD}, target localization \cite{Sun_TL,Zhang_TL},
etc. For a standard compressed sensing problem, a sparse signal $\boldsymbol{x}\in\mathbb{C}^{N\times1}$
is to be recovered from measurements $\boldsymbol{y}\in\mathbb{C}^{M\times1}$
($M<N$) under a linear observation model,
\begin{equation}
\boldsymbol{\boldsymbol{y}}=\mathbf{F}\boldsymbol{x}+\boldsymbol{w},\label{eq:strandard model}
\end{equation}
where the sensing matrix $\mathbf{F}\in\mathbb{C}^{M\times N}$ is
fixed and perfectly known, and the noise vector $\boldsymbol{w}\in\mathbb{C}^{M\times1}$
follows a complex Gaussian distribution with noise variance $\gamma^{-1}$.
However, in many practical scenarios, some dynamic grid parameters
may exist in the sensing matrix. For instance, in massive multiple-input
multiple-output (MIMO) systems, the angular-domain dynamic gird parameters
are usually introduced for high-performance channel estimation \cite{Dai_MIMOCE}.
In this case, the observation model in $(\ref{eq:strandard model})$
can be rewritten into
\begin{equation}
\boldsymbol{\boldsymbol{y}}=\mathbf{F}\left(\boldsymbol{\theta}\right)\boldsymbol{x}+\boldsymbol{w},\label{eq:off-grid model}
\end{equation}
where $\boldsymbol{\theta}$ denotes the dynamic grid parameters.
Our primary goal is to recover the sparse signal $\boldsymbol{x}$
and estimate the dynamic grid parameters $\boldsymbol{\theta}$ simultaneously
given observations $\boldsymbol{y}$. There are three common methods
in the literature.

\textbf{On-grid based CS methods:} The main idea of the on-grid based
method is to select a fixed sampling grid and use discrete grid points
to approximate the true parameters $\boldsymbol{\theta}$. The conventional
CS method is a good choice under the on-grid based model, such as
orthogonal matching pursuit (OMP) \cite{Tropp_CE_OMP}, $\ell_{1}$-norm
optimization \cite{Malioutov_l1,LiuAn_CE_Burst_LASSO}, and sparse
Bayesian learning/inference \cite{Tipping_SBL,Ji_SBL,Babacan_SBL}.
In practice, the true parameters $\boldsymbol{\theta}$ usually do
not lie exactly on the fixed grid points. And thus the estimation
accuracy of $\boldsymbol{\theta}$ is limited by the grid resolution.
To reduce the mismatch between the true parameter and its nearest
grid point, a dense sampling grid is needed. However, a dense sampling
grid leads to a highly correlated sensing matrix and a poor estimate
of the sparse signal.

\textbf{Off-grid sparse Bayesian inference (OGSBI):}\textit{ }It is
very challenging to directly estimate $\boldsymbol{\theta}$ since
the mapping $\boldsymbol{\theta}\rightarrow\mathbf{F}\left(\boldsymbol{\theta}\right)$
is nonlinear. To address this difficulty, the authors in \textcolor{blue}{\cite{OGSBI,ChengLei_Bean_Squint}}
approximated the basis vectors of the sensing matrix using linearization.
The proposed OGSBI algorithm achieved a better performance than the
on-grid based CS methods. However, the error caused by linear approximation
is not completely eliminated due to the absence of high-order items
of Taylor expansion. In \cite{W-OGSBI}, the authors improved the
OGSBI algorithm and proposed a new weighted OGSBI algorithm based
on second-order Taylor expansion approximation. Another main drawback
of the OGSBI is that the Laplace prior model used in \cite{OGSBI,ChengLei_Bean_Squint,W-OGSBI}
can only exploit an i.i.d. sparse structure.

\textbf{Expectation maximization (EM)-based methods:} The EM-based
methods contain two major steps, where the E-step computes a Bayesian
estimation of $\boldsymbol{x}$ and the M-step gives a point estimation
of $\boldsymbol{\theta}$. To describe different type of sparse structures,
some recent literature usually adopted the turbo approach as the E-step.
In \cite{Som_TurboAMP}, the authors proposed a novel turbo approximate
message passing (Turbo-AMP) for loopy belief propagation. Inspired
by this work, the E-step in \cite{Yuan_TurboCS,LiuAn_TurboOAMP,Huangzhe_TurboSBI}
was a turbo compressed sensing (Turbo-CS) framework, by combining
the linear minimum mean square error (LMMSE) estimator with message
passing. In \cite{LiuAn_CE_Turbo_VBI,LiuAn_directloc_vehicles}, the
authors proposed a turbo variational Bayesian inference (Turbo-VBI)
algorithm by combining the VBI estimator with message passing. The
EM-based methods can exploit more complicated sparse structures and
achieve better performance than the first two methods. However, the
computational complexity of these methods are often higher. In the
first place, the Bayesian estimator in the E-step usually involves
a matrix inverse in each iteration. Although it is possible to avoid
matrix inverse for a few special choices of sensing matrix (such as
Turbo-AMP for an i.i.d. sensing matrix and Turbo-CS for a partially
orthogonal sensing matrix), the sensing matrices in many important
practical applications do not belong to these special cases, especially
under the consideration of dynamic grid parameters. Secondly, the
EM-based methods involve double-loop iterations, i.e., the inner iteration
involved in the E-step (Bayesian estimator) and the outer iteration
between the E-step and M-step.

In this paper, we propose an inverse-free successive linear approximation
VBI algorithm to overcome the drawbacks of the existing methods. The
proposed algorithm can output the Bayesian estimation of both sparse
signals and dynamic grid parameters, and it can achieve lower complexity,
faster convergence, and better performance compared to the state-of-the-art
EM-based Turbo-CS and Turbo-VBI algorithms. The main contributions
are summarized below.
\begin{itemize}
\item \textbf{Successive linear approximation VBI (SLA-VBI):} We aim at
computing the approximate posterior distribution of both sparse signals
and dynamic grid parameters based on the VBI iterations. Using the
successive linear approximation approach, the Bayesian inference can
be performed in closed form. In contrast to the double-loop EM-based
methods, the proposed SLA-VBI reduces the number of iterations significantly
while speeding up convergence.
\item \textbf{Inverse-free algorithm design:} Conventional sparse Bayesian
inference algorithms usually involve the matrix inverse operation
during iterations. To reduce the computational overhead, we adopt
the majorization-minimization (MM) \cite{Sun_MM} framework to avoid
the matrix inverse and propose a low-complexity inverse-free successive
linear approximation VBI (IFSLA-VBI) algorithm. The proposed IFSLA-VBI
can achieve a better trade-off between performance and complexity
by controlling the number of iterations used to approximate the matrix
inverse according to the structure of the sensing matrix.
\item \textbf{Extension to structured sparse priors for practical applications:
}In practical applications, the sparse signal $\boldsymbol{x}$ usually
has structured sparsity. To exploit the specific sparse structures,
we extend our proposed algorithm from an independent sparse prior
to more complicated structured sparse priors and apply it to solve
important practical problems in wireless communications.
\end{itemize}
The rest of the paper is organized as follows. In Section II, we introduce
a three-layer sparse prior model and present the system model of two
practical applications. In Section III, we introduce the proposed
SLA-VBI and IFSLA-VBI algorithms. In Section IV, we elaborate on how
to extend the proposed algorithm to structured sparse priors. Simulation
results and conclusions are shown in Section V and VI, respectively.

\textit{Notation:} Lowercase boldface letters denote vectors and uppercase
boldface letters denote matrices. $\left(\cdot\right)^{-1}$, $\left(\cdot\right)^{T}$,
$\left(\cdot\right)^{H}$, $\left\Vert \cdot\right\Vert $, $\left\langle \cdot\right\rangle $,
and $\textrm{diag}\left(\cdot\right)$ are used to represent the inverse,
transpose, conjugate transpose, $\ell_{2}\textrm{-norm}$, expectation,
and diagonalization operations, respectively. Let $\mathfrak{Re}\left\{ \cdot\right\} $
denote the real part of the complex argument. For a set $\mathcal{N}$,
we use $\left|\mathcal{N}\right|$ to denote its cardinality. Let
$\boldsymbol{x}\triangleq\left[x_{n}\right]_{n\in\mathcal{N}}\in\mathbb{C}^{\left|\mathcal{N}\right|\times1}$
represent a vector composed of elements indexed by $\mathcal{N}$.
$\mathcal{CN}\left(\boldsymbol{x};\boldsymbol{\mu},\mathbf{\Sigma}\right)$
represents a complex Gaussian distribution with mean $\boldsymbol{\mu}$
and covariance matrix $\mathbf{\Sigma}$. $\textrm{Ga}\left(x;a,b\right)$
represents a Gamma distribution with shape parameter $a$ and rate
parameter $b$.

\section{System Model}

\subsection{Three-layer Sparse Prior Model}

\begin{figure}[t]
\begin{centering}
\includegraphics[width=80mm]{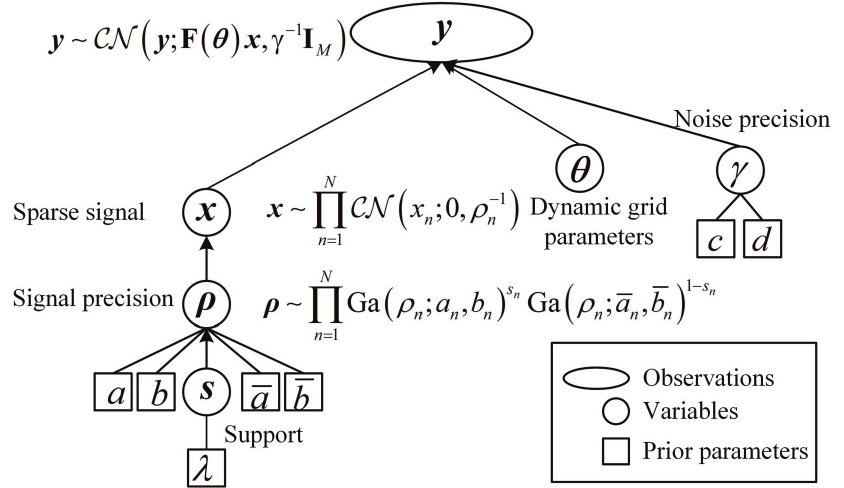}
\par\end{centering}
\caption{\label{fig:Sparse_prior_model}Illustration of the three-layer hierarchical
sparse prior model.}
\end{figure}
We introduce a three-layer sparse prior model \cite{LiuAn_CE_Turbo_VBI,LiuAn_directloc_vehicles}
that can describe various sparse structures, as illustrated in Fig.
\ref{fig:Sparse_prior_model}. Specifically, we use a binary vector
$\boldsymbol{s}\triangleq\left[s_{1},\ldots,s_{N}\right]^{T}$ to
represent the support of $\boldsymbol{x}$, where $s_{n}=1$ indicates
$x_{n}$ is non-zero and $s_{n}=0$ indicates the opposite. Let $\boldsymbol{\rho}\triangleq\left[\rho_{1},\ldots,\rho_{N}\right]^{T}$
denote the precision vector of $\boldsymbol{x}$, where $1/\rho_{n}$
is the variance of $x_{n}$. The joint distribution of $\boldsymbol{x}$,
$\boldsymbol{\rho}$, and $\boldsymbol{s}$ can be expressed as
\begin{equation}
p\left(\boldsymbol{x},\boldsymbol{\rho},\boldsymbol{s}\right)=\underbrace{p\left(\boldsymbol{s}\right)}_{\textrm{Support}}\underbrace{p\left(\boldsymbol{\rho}\mid\boldsymbol{s}\right)}_{\textrm{Precision}}\underbrace{p\left(\boldsymbol{x}\mid\boldsymbol{\rho}\right)}_{\textrm{Sparse\ signal}}.\label{eq:p(x,rou,s)}
\end{equation}
A complex Gaussian distribution is assumed as the prior for $\boldsymbol{x}$.
Moreover, conditioned on $\boldsymbol{\rho}$, the elements of $\boldsymbol{x}$
are independent, i.e.,
\begin{equation}
p\left(\boldsymbol{x}\mid\boldsymbol{\rho}\right)=\prod_{n=1}^{N}p\left(x_{n}\mid\rho_{n}\right)=\prod_{n=1}^{N}\mathcal{CN}\left(x_{n};0,\rho_{n}^{-1}\right).
\end{equation}
The precision vector $\boldsymbol{\rho}$ can be expressed with the
Bernoulli-Gamma distribution
\begin{align}
p\left(\boldsymbol{\rho}\mid\boldsymbol{s}\right)= & \prod_{n=1}^{N}\textrm{Ga}\left(\rho_{n};a_{n},b_{n}\right)^{s_{n}}\textrm{Ga}\left(\rho_{n};\overline{a}_{n},\overline{b}_{n}\right)^{1-s_{n}},
\end{align}
where $a_{n}$, $b_{n}$ and $\overline{a}_{n}$, $\overline{b}_{n}$
are prior parameters of $\rho_{n}$ conditioned on $s_{n}=1$ and
$s_{n}=0$, respectively. To indicate $x_{n}$ is zero or non-zero
more effectively, $a_{n}$ and $b_{n}$ are chosen to satisfy $\frac{a_{n}}{b_{n}}=\mathbb{E}\left(\rho_{n}\mid s_{n}=1\right)=\Theta\left(1\right)$,
while $\overline{a}_{n}$ and $\overline{b}_{n}$ are chosen to satisfy
$\frac{a_{n}}{b_{n}}=\mathbb{E}\left(\rho_{n}\mid s_{n}=0\right)\gg1$
\cite{LiuAn_CE_Turbo_VBI,LiuAn_directloc_vehicles}.

\textcolor{blue}{The prior for the support vector depends on the specific
sparse structure. For example, for an independent sparse structure,
a Bernoulli distribution is usually used as the prior,}
\begin{equation}
p\left(\boldsymbol{s}\right)=\prod_{n=1}^{N}\left(\lambda_{n}\right)^{s_{n}}\left(1-\lambda_{n}\right)^{1-s_{n}},
\end{equation}
where $\lambda_{n}$ gives the probability of $p\left(s_{n}=1\right)$.
\textcolor{blue}{For more complicated sparse structures, we use other
sparse priors to capture the specific structured sparsity. }And our
proposed algorithm can be easily extended to these cases via the turbo
approach.\textcolor{blue}{{} }We will elaborate on the extended algorithm
in Section \ref{sec:Extension-to-Structured}.

Meanwhile, we employ a gamma distribution with parameters $c$ and
$d$ to model the noise precision, i.e.,
\begin{equation}
p\left(\gamma\right)=\textrm{Ga}\left(\gamma;c,d\right).\label{eq:p(gamma)}
\end{equation}

\subsection{Problem Statement}

Recall the linear observation model with dynamic grid parameters in
the sensing matrix
\begin{equation}
\boldsymbol{\boldsymbol{y}}=\mathbf{F}\left(\boldsymbol{\theta}\right)\boldsymbol{x}+\boldsymbol{w}.\label{eq:final model}
\end{equation}
According to the physical meaning, we partition $\boldsymbol{\theta}$
into $B$ blocks $\boldsymbol{\theta}\triangleq\bigl\{\boldsymbol{\theta}^{1},\ldots,\boldsymbol{\theta}^{B}\bigr\}$
, such that each block $\boldsymbol{\theta}^{j}\triangleq\bigl[\theta_{1}^{j},\ldots,\theta_{N}^{j}\bigr]^{T}\in\mathbb{C}^{N\times1},j\in\left\{ 1,\ldots,B\right\} $
denotes a type of dynamic grid parameters. For example, we partition
dynamic grid parameters into distance parameters $\boldsymbol{\iota}$
and angle parameters $\boldsymbol{\vartheta}$ in subsection \ref{subsec:5G-based-Target-Detection}.
Let $\mathbf{F}\left(\boldsymbol{\theta}\right)\triangleq\left[\boldsymbol{\varPhi}\left(\boldsymbol{\theta}_{1}\right),\ldots,\boldsymbol{\varPhi}\left(\boldsymbol{\theta}_{N}\right)\right]$,
where $\boldsymbol{\theta}_{n}\triangleq\bigl[\theta_{n}^{1},\ldots,\theta_{n}^{B}\bigr]^{T}$
denotes parameters of the $n\textrm{-th}$ basis vector $\boldsymbol{\varPhi}\left(\boldsymbol{\theta}_{n}\right)$
for $n=1,\ldots,N$. Our primary goal is to compute the Bayesian estimation
of the sparse signal $\boldsymbol{x}$, the support vector $\boldsymbol{s}$,
and the dynamic grid parameters $\boldsymbol{\theta}$ given the observations
$\boldsymbol{y}$. Such a joint sparse signal recovery and dynamic
grid estimation problem includes many important application problems
as special cases. In the following two subsections, we shall present
two application examples.

\subsection{Massive MIMO Channel Estimation with Limited Pilots\label{subsec:Massive-MIMO-Channel}}

\begin{figure}[t]
\begin{centering}
\includegraphics[width=80mm]{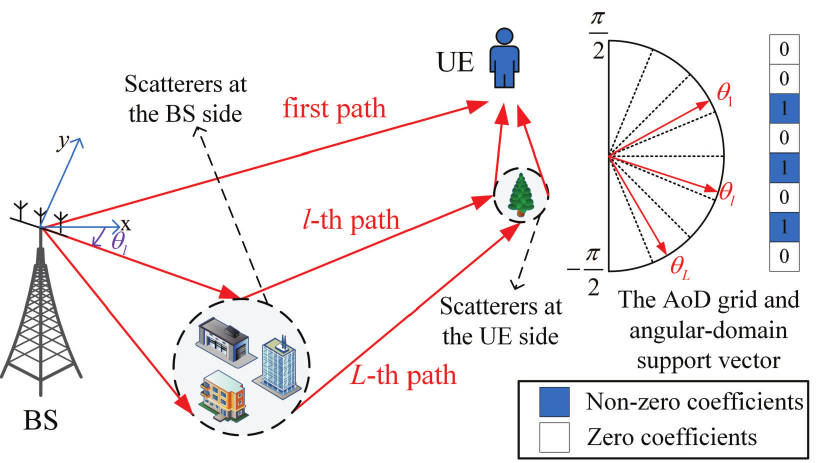}
\par\end{centering}
\textcolor{blue}{\caption{\label{fig:FDD_CE}\textcolor{blue}{Illustration of the massive MIMO
downlink channel and its non-zero coefficients.}}
}
\end{figure}

Consider a narrow-band massive MIMO system with a base station (BS)
serving a single-antenna user, as shown in Fig. \ref{fig:FDD_CE}.
The BS is equipped with a uniform linear array (ULA) of $N\gg1$ antennas.
To estimate the downlink channel vector $\boldsymbol{h}\in\mathbb{C}^{N\times1}$,
the BS transmits pilot sequences $\boldsymbol{u}_{t}\in\mathbb{C}^{N\times1},t=1,\ldots,M$
($M<N$) to the user. The received signal $\boldsymbol{y}\in\mathbb{C}^{M\times1}$
can be expressed as
\begin{equation}
\boldsymbol{y}=\mathbf{U}\boldsymbol{h}+\boldsymbol{w},\label{eq:FFD_y}
\end{equation}
where $\mathbf{U}\triangleq\left[\boldsymbol{u}_{1},\ldots,\boldsymbol{u}_{M}\right]^{T}\in\mathbb{C}^{M\times N}$
denotes the pilot matrix and $\boldsymbol{w}\sim\mathcal{CN}\left(\boldsymbol{w};0,\gamma^{-1}\mathbf{I}_{M}\right)$
is the Gaussian noise. Assume there are $L$ paths for the communication
channel, the channel vector $\boldsymbol{h}$ can be modeled as
\begin{equation}
\boldsymbol{h}=\sum_{l=1}^{L}x_{l}\boldsymbol{a}\left(\theta_{l}\right),\label{eq:FDD_h}
\end{equation}
where $x_{l}$ and $\theta_{l}$ denote the complex channel gain and
angle-of-departure (AoD) of the $l\textrm{-th}$ path, respectively.
The steer vector at the BS is given by
\begin{equation}
\boldsymbol{a}\bigl(\theta\bigr)\triangleq\frac{1}{\sqrt{N}}\left[1,e^{j\pi\sin\theta},\ldots,e^{j\left(N-1\right)\pi\sin\theta}\right]^{T}.
\end{equation}
To obtain a sparse representation of the channel vector, we adopt
the grid-based solution. Specifically, we define a fixed grid $\left\{ \overline{\vartheta}_{1},\ldots,\overline{\vartheta}_{\widetilde{N}}\right\} $
of $\widetilde{N}$ AoD points such that $\left\{ \sin\overline{\vartheta}_{n}\right\} _{n=1}^{_{\widetilde{N}}}$
are uniformly distributed in the range $\left[-1,1\right]$. 

However, the true AoDs usually do not lie exactly on $\widetilde{N}$
discrete AoD grid points. In this case, the gap between the true AoD
and its nearest grid point will lead to energy leakage. To mitigate
the effect of energy leakage, we introduce a dynamic AoD grid $\boldsymbol{\vartheta}=\left[\vartheta_{1},\ldots,\vartheta{}_{\widetilde{N}}\right]^{T}$
instead of only using a fixed sampling grid \footnote{The fixed grid $\left\{ \overline{\vartheta}_{1},\ldots,\overline{\vartheta}_{\widetilde{N}}\right\} $
is usually chosen as the initial value of the dynamic grid in the
algorithm.}. In the algorithm design, the grid parameters $\boldsymbol{\vartheta}$
will be updated via Bayesian inference for more accurate channel estimation.

Based on the definition of the dynamic AoD grid, we can obtain a sparse
basis $\mathbf{A}\left(\boldsymbol{\vartheta}\right)\triangleq\left[\boldsymbol{a}\left(\vartheta_{1}\right),\ldots,\boldsymbol{a}\left(\vartheta{}_{\widetilde{N}}\right)\right]\in\mathbb{C}^{N\times\widetilde{N}}$.
The sparse representation of the channel vector in (\ref{eq:FDD_h})
is given by
\begin{equation}
\boldsymbol{h}=\mathbf{A}\left(\boldsymbol{\vartheta}\right)\boldsymbol{x},\label{eq:FDD_sparse_h}
\end{equation}
where $\boldsymbol{x}\in\mathbb{C}^{\widetilde{N}\times1}$ is the
angular-domain sparse channel vector. $\boldsymbol{x}$ has only $L$
non-zero elements corresponding to the AoDs of $L\ll\widetilde{N}$
paths. \textcolor{blue}{Let $\boldsymbol{s}\triangleq\bigl[s_{1},\ldots,s_{_{\widetilde{N}}}\bigr]^{T}$
denote the support vector of $\boldsymbol{x}$, where $s_{n}=1$ indicates
there is a channel path with AoD $\vartheta_{n}$, while $s_{n}=0$
indicates the opposite.}

Then the received signal in (\ref{eq:FFD_y}) can be rewritten into
\begin{equation}
\boldsymbol{y}=\mathbf{F}\left(\boldsymbol{\vartheta}\right)\boldsymbol{x}+\boldsymbol{w},
\end{equation}
where $\mathbf{F}\left(\boldsymbol{\vartheta}\right)\triangleq\mathbf{U}\mathbf{A}\left(\boldsymbol{\vartheta}\right)$. 

For such a linear observation model with dynamic grid parameters in
the sensing matrix, our goal is to recover the angular-domain sparse
channel vector $\boldsymbol{x}$, the support vector $\boldsymbol{s}$,
and the AoD grid parameters $\boldsymbol{\vartheta}$ from the received
signal $\boldsymbol{y}$.

\subsection{6G-based Target Detection and Localization\label{subsec:5G-based-Target-Detection}}

\begin{figure}[t]
\begin{centering}
\includegraphics[width=80mm]{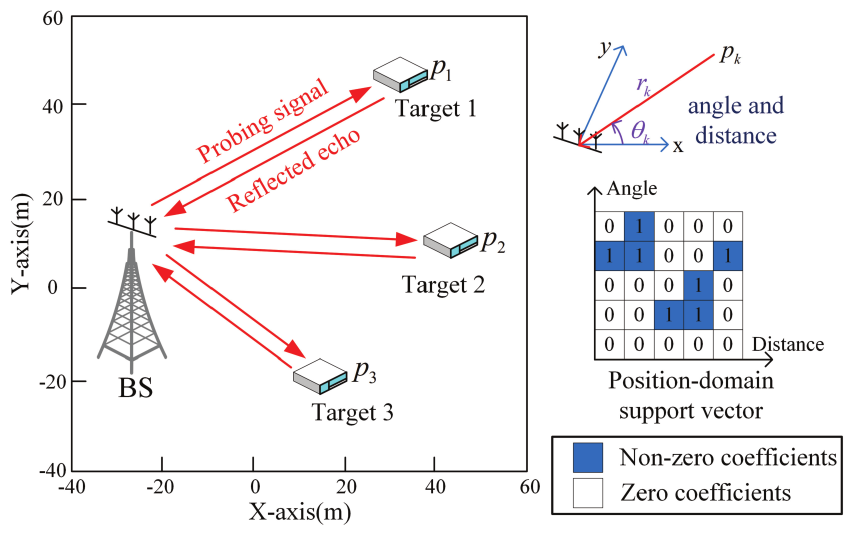}
\par\end{centering}
\caption{\label{fig:target detection}A target detection and localization model
in 6G MIMO-OFDM systems and the non-zero coefficients of the virtual
position-domain channel.}
\end{figure}

Consider a broadband MIMO Orthogonal Frequency Division Multiplexing
(OFDM) system with a BS equipped with $N$ antennas and $N_{\textrm{RF}}<N$
radio frequency (RF) chains, as illustrated in Fig. \ref{fig:target detection}.
In future 6G wireless systems, the MIMO-OFDM signal will also be exploited
to provide target sensing functionality \cite{Zhang_MIMOOFDM_Radar,Xu_MIMOOFDM_Radar}.
For simplicity, we consider that all targets are on a two-dimensional
(2-D) plane. Note that our results can be easily extended to a 3-D
environment. We assume there are $K$ targets in the area, and the
polar coordinates of the $k\textrm{-th}$ target is represented as
$\boldsymbol{p}_{k}\triangleq\left(r_{k},\theta_{k}\right)$, where
$r_{k}$ is its distance from the BS and $\theta_{k}$ is its angle.

To sense the presence of the targets and estimate the associated parameters,
on the $n\textrm{-th}$ subcarrier for $n\in\mathcal{N}_{b}$, \textcolor{blue}{the
BS sends a probing signal $\boldsymbol{u}_{n}\in\mathbb{C}^{N\times1}$}.
Then the reflected echo signal $\boldsymbol{\boldsymbol{y}}_{n}\in\mathbb{C}^{N_{\textrm{RF}}\times1}$
can be expressed as
\begin{equation}
\boldsymbol{\boldsymbol{y}}_{n}=\mathbf{W}_{\textrm{RF}}\mathbf{H}_{n}\boldsymbol{u}_{n}+\boldsymbol{w}_{n},\forall n\in\mathcal{N}_{b},\label{eq:yn}
\end{equation}
where $\mathbf{W}_{\textrm{RF}}\in\mathbb{C}^{N_{\textrm{RF}}\times N}$
is the RF combining matrix, $\boldsymbol{w}_{n}\sim\mathcal{CN}\left(\boldsymbol{w}_{n};0,\gamma^{-1}\mathbf{I}_{N_{\textrm{RF}}}\right)$
is the noise vector, and the channel matrix $\mathbf{H}_{n}\in\mathbb{C}^{N\times N}$
is modeled as
\begin{equation}
\mathbf{H}_{n}=\sum_{k=1}^{K}x_{k}e^{-j2\pi\left(n-1\right)f_{0}\tau_{k}}\boldsymbol{a}\left(\theta_{k}\right)\boldsymbol{a}^{T}\left(\theta_{k}\right),
\end{equation}
where $x_{k}$ is the complex reflection coefficient of the $k\textrm{-th}$
target and $f_{0}$ is the subcarrier interval. The propagation delay
is denoted by $\tau_{k}=2r_{k}/c$, where $c$ is the speed of light.

Similar to subsection \ref{subsec:Massive-MIMO-Channel}, we introduce
a dynamic position grid $\left\{ \boldsymbol{\iota},\boldsymbol{\vartheta}\right\} \triangleq\left\{ \left(\iota_{1},\vartheta_{1}\right),\ldots,\left(\iota_{Q},\vartheta_{Q}\right)\right\} $
of $Q$ position grid points for high-accuracy target localization,
where $\boldsymbol{\iota}$ and $\boldsymbol{\vartheta}$ denote distance
and angle parameters, respectively.

With the definition of the dynamic position grid, we define a sparse
basis as 
\[
\mathbf{B}_{n}\left(\boldsymbol{\iota},\boldsymbol{\vartheta}\right)\triangleq\left[\boldsymbol{b}_{n,1},\ldots,\boldsymbol{b}_{n,Q}\right]\in\mathbb{C}^{N\times Q},\forall n\in\mathcal{N}_{b},
\]
where the basis vector is given by 
\[
\boldsymbol{b}_{n,q}\triangleq e^{-j2\pi\left(n-1\right)f_{0}\frac{2\iota_{q}}{c}}\boldsymbol{a}\left(\vartheta_{q}\right)\boldsymbol{a}^{T}\left(\vartheta_{q}\right)\boldsymbol{u}_{n},\forall n\in\mathcal{N}_{b},\forall q.
\]
Then the echo signal in (\ref{eq:yn}) can be rewritten into
\begin{equation}
\boldsymbol{\boldsymbol{y}}_{n}=\mathbf{W}_{\textrm{RF}}\mathbf{B}_{n}\left(\boldsymbol{\iota},\boldsymbol{\vartheta}\right)\boldsymbol{x}+\boldsymbol{w}_{n},\forall n\in\mathcal{N}_{b},\label{eq:yn_sparse}
\end{equation}
where $\boldsymbol{x}\triangleq\left[x_{1},\ldots,x_{Q}\right]^{T}\in\mathbb{C}^{Q\times1}$
is called the position-domain sparse channel vector. We use $\boldsymbol{s}\triangleq\left[s_{1},\ldots,s_{Q}\right]^{T}$
to represent the support vector of $\boldsymbol{x}$, where $s_{q}=1$
indicates there is a target lying in the $q\textrm{-th}$ position
grid with angle $\vartheta_{q}$ and distance $\iota_{q}$, while
$s_{q}=0$ indicates the opposite.

Using (\ref{eq:yn_sparse}), the received echo signal on all available
subcarriers can be obtained as
\begin{equation}
\boldsymbol{\boldsymbol{y}}=\mathbf{F}\left(\boldsymbol{\iota},\boldsymbol{\vartheta}\right)\boldsymbol{x}+\boldsymbol{w},
\end{equation}
where $\boldsymbol{\boldsymbol{y}}\triangleq\left[\boldsymbol{y}_{n}\right]_{n\in\mathcal{N}_{b}}$,
$\mathbf{F}\left(\boldsymbol{\iota},\boldsymbol{\vartheta}\right)\triangleq\left[\mathbf{W}_{\textrm{RF}}\mathbf{B}_{n}\bigl(\boldsymbol{\iota},\boldsymbol{\vartheta}\bigr)\right]_{_{n\in\mathcal{N}_{b}}}$,
and $\boldsymbol{w}\triangleq\left[\boldsymbol{w}_{n}\right]_{n\in\mathcal{N}_{b}}$. 

For a target detection and localization problem, we aim at estimating
the support vector $\boldsymbol{s}$ and the position grid parameters
$\left\{ \boldsymbol{\iota},\boldsymbol{\vartheta}\right\} $ from
the received echo signal $\boldsymbol{y}$.

\section{Successive Linear Approximation VBI}

\subsection{Mean Field VBI}

We first give an overview of the mean field variational Bayesian inference.
Let $\boldsymbol{v}\triangleq\left\{ \boldsymbol{x},\boldsymbol{\rho},\boldsymbol{s},\gamma,\boldsymbol{\theta}\right\} $
denote the collection of hidden variables in (\ref{eq:final model}).
\textcolor{blue}{For convenience, we use $\boldsymbol{v}^{l}$ to
denote an individual variable in $\boldsymbol{v}$ and let $\mathcal{H}\triangleq\left\{ l\mid\forall\boldsymbol{v}^{l}\in\boldsymbol{v}\right\} $.}
We aim at calculating the posterior distribution of hidden variables,
i.e., $p\left(\boldsymbol{v}\mid\boldsymbol{y}\right)$. However,
it is usually intractable to find the posterior directly since the
considered problem involves integrals of many high-dimensional variables.
\textcolor{blue}{Based on the mean field VBI method, the posterior
distribution is approximated by the variational distribution $q\left(\boldsymbol{v}\right)$
that minimizes the Kullback-Leibler (KL) divergence between $q\left(\boldsymbol{v}\right)$
and $p\left(\boldsymbol{v}\mid\boldsymbol{y}\right)$ under a factorized
form constraint as
\begin{equation}
\begin{aligned} & \min_{q\left(\boldsymbol{v}\right)} & \mathrm{KL}\left(q\bigparallel p\right) & \triangleq\int q\left(\boldsymbol{v}\right)\ln\frac{q\left(\boldsymbol{v}\right)}{p\left(\boldsymbol{v}\mid\boldsymbol{\boldsymbol{y}}\right)}\mathrm{d}\boldsymbol{v},\\
 & s.t. & q\left(\boldsymbol{v}\right) & =\prod_{l\in\mathcal{H}}q\left(\boldsymbol{v}^{l}\right),
\end{aligned}
\label{eq:problem_VBI}
\end{equation}
where the constraint $q\left(\boldsymbol{v}\right)=\prod_{l\in\mathcal{H}}q\left(\boldsymbol{v}^{l}\right)$
is the mean field assumption \cite{Parisi_MeanField}. Note that the
dynamic grid $\boldsymbol{\theta}$ is also viewed as a hidden variable
in algorithm design, which is quite different from the conventional
VBI in \cite{LiuAn_CE_Turbo_VBI,LiuAn_directloc_vehicles}. And thus,
our proposed algorithm can provide the Bayesian estimation of $\boldsymbol{\theta}$
additionally.}

\textcolor{blue}{Although the problem (\ref{eq:problem_VBI}) is known
to be non-convex, it is convex w.r.t a single variational distribution
$q\left(\boldsymbol{v}^{l}\right)$ after fixing other variational
distributions $q\left(\boldsymbol{v}^{k}\right),\forall k\neq l$
\cite{Tzikas_VBI}. And it has been proved in \cite{Tzikas_VBI} that
a stationary solution could be found via optimizing each variational
distribution in an alternating fashion.} \textcolor{blue}{Specifically,
for given $q\left(\boldsymbol{v}^{k}\right),\forall k\neq l$, the
optimal $q\left(\boldsymbol{v}^{l}\right)$ that minimizes the KL-divergence
is given by \cite{Tzikas_VBI}
\begin{equation}
q\bigl(\boldsymbol{v}^{l}\bigr)=\frac{\exp\left(\left\langle \ln p\left(\boldsymbol{v},\boldsymbol{y}\right)\right\rangle _{\Pi_{k\neq l}q\left(\boldsymbol{v}^{k}\right)}\right)}{\int\exp\left(\left\langle \ln p\left(\boldsymbol{v},\boldsymbol{y}\right)\right\rangle _{\Pi_{k\neq l}q\left(\boldsymbol{v}^{k}\right)}\right)\textrm{d}\boldsymbol{v}^{l}},\label{eq:q(vl)}
\end{equation}
where $\left\langle \cdot\right\rangle _{\Pi_{k\neq l}q\left(\boldsymbol{v}^{k}\right)}$
is an expectation operation w.r.t. $q\left(\boldsymbol{v}^{k}\right)$
for $k\neq l$. The joint distribution $p\left(\boldsymbol{v},\boldsymbol{y}\right)$
is given by
\begin{align}
p\left(\boldsymbol{v},\boldsymbol{y}\right) & =p\left(\boldsymbol{y}\mid\boldsymbol{x},\boldsymbol{\theta},\gamma\right)p\left(\boldsymbol{x},\boldsymbol{\rho},\boldsymbol{s}\right)p\left(\gamma\right)p\left(\boldsymbol{\theta}\right),\label{eq:joint distribution}
\end{align}
where $p\left(\boldsymbol{y}\mid\boldsymbol{x},\boldsymbol{\theta},\gamma\right)=\mathcal{CN}\left(\boldsymbol{y};\mathbf{F}\left(\boldsymbol{\theta}\right)\boldsymbol{x},\gamma^{-1}\mathbf{I}_{M}\right)$
is the likelihood function, $p\left(\boldsymbol{x},\boldsymbol{\rho},\boldsymbol{s}\right)$
and $p\left(\gamma\right)$ are the priors given in (\ref{eq:p(x,rou,s)})
and (\ref{eq:p(gamma)}), respectively, and $p\left(\boldsymbol{\theta}\right)$
is the prior for the dynamic grid $\boldsymbol{\theta}$.}

\textcolor{blue}{By substituting the joint distribution (\ref{eq:joint distribution})
into (\ref{eq:q(vl)}), each optimal variational distribution $q\bigl(\boldsymbol{v}^{l}\bigr)$
can be derived. In the following, we shall provide the details of
the derivation for each variational distribution.}

\subsubsection{\textcolor{blue}{Update of $q\left(\boldsymbol{x}\right)$}}

\textcolor{blue}{Using (\ref{eq:q(vl)}) and ignoring the terms that
are not related to $\boldsymbol{x}$, the posterior distribution $q\left(\boldsymbol{x}\right)$
can be derived as
\begin{align}
\ln q\left(\boldsymbol{x}\right)\propto & \left\langle \ln p\left(\boldsymbol{v},\boldsymbol{y}\right)\right\rangle _{q\left(\boldsymbol{\rho}\right)q\left(\boldsymbol{s}\right)q\left(\gamma\right)q\left(\boldsymbol{\theta}\right)}\nonumber \\
\propto & \left\langle \ln p\left(\boldsymbol{y}\mid\boldsymbol{x},\boldsymbol{\theta},\gamma\right)\right\rangle _{q\left(\gamma\right)q\left(\boldsymbol{\theta}\right)}+\left\langle \ln p\left(\boldsymbol{x}\mid\boldsymbol{\rho}\right)\right\rangle _{q\left(\boldsymbol{\rho}\right)}\nonumber \\
\propto & -\bigl\langle\gamma\bigr\rangle\left\langle \left\Vert \boldsymbol{y}-\mathbf{F}\left(\boldsymbol{\theta}\right)\boldsymbol{x}\right\Vert ^{2}\right\rangle _{q\left(\boldsymbol{\theta}\right)}-\boldsymbol{x}^{H}\textrm{diag}\left(\left\langle \boldsymbol{\rho}\right\rangle \right)\boldsymbol{x}\nonumber \\
\propto & -\boldsymbol{x}^{H}\left(\left\langle \gamma\right\rangle \left\langle \mathbf{F}\left(\boldsymbol{\theta}\right)^{H}\mathbf{F}\left(\boldsymbol{\theta}\right)\right\rangle _{q\left(\boldsymbol{\theta}\right)}+\textrm{diag}\left(\left\langle \boldsymbol{\rho}\right\rangle \right)\right)\boldsymbol{x}\nonumber \\
 & +2\mathfrak{Re}\left\{ \boldsymbol{x}^{H}\left\langle \gamma\right\rangle \left\langle \mathbf{F}\left(\boldsymbol{\theta}\right)^{H}\right\rangle _{q\left(\boldsymbol{\theta}\right)}\boldsymbol{y}\right\} \nonumber \\
\propto & -\boldsymbol{x}^{H}\boldsymbol{\Sigma}_{x}^{-1}\boldsymbol{x}+2\mathfrak{Re}\left\{ \boldsymbol{x}^{H}\boldsymbol{\Sigma}_{x}^{-1}\boldsymbol{\mu}_{x}\right\} .\label{eq:lnq(x)}
\end{align}
Clearly, this is the exponent of a complex Gaussian distribution with
mean $\boldsymbol{\mu}_{x}$ and covariance matrix $\boldsymbol{\Sigma}_{x}$
given by
\begin{equation}
\begin{aligned}\boldsymbol{\mu}_{x} & =\boldsymbol{\Sigma}_{x}\left\langle \gamma\right\rangle \left\langle \mathbf{F}\left(\boldsymbol{\theta}\right)^{H}\right\rangle _{q\left(\boldsymbol{\theta}\right)}\boldsymbol{y},\\
\boldsymbol{\Sigma}_{x} & =\left(\left\langle \gamma\right\rangle \left\langle \mathbf{F}\left(\boldsymbol{\theta}\right)^{H}\mathbf{F}\left(\boldsymbol{\theta}\right)\right\rangle _{q\left(\boldsymbol{\theta}\right)}+\textrm{diag}\left(\left\langle \boldsymbol{\rho}\right\rangle \right)\right)^{-1}.
\end{aligned}
\label{eq:x_post_old}
\end{equation}
}

\subsubsection{\textcolor{blue}{Update of $q\left(\boldsymbol{\rho}\right)$}}

\textcolor{blue}{The posterior distribution $q\left(\boldsymbol{\rho}\right)$
can be computed by
\begin{align}
\ln q\left(\boldsymbol{\rho}\right)\propto & \left\langle \ln p\left(\boldsymbol{v},\boldsymbol{y}\right)\right\rangle _{q\left(\boldsymbol{x}\right)q\left(\boldsymbol{s}\right)q\left(\gamma\right)q\left(\boldsymbol{\theta}\right)}\nonumber \\
\propto & \left\langle \ln p\left(\boldsymbol{x}\mid\boldsymbol{\rho}\right)\right\rangle _{q\left(\boldsymbol{x}\right)}+\left\langle \ln p\left(\boldsymbol{\rho}\mid\boldsymbol{s}\right)\right\rangle _{q\left(\boldsymbol{s}\right)}\nonumber \\
\propto & \sum_{n=1}^{N}\ln\rho_{n}-\sum_{n=1}^{N}\rho_{n}\left\langle x_{n}^{2}\right\rangle \nonumber \\
 & +\sum_{n=1}^{N}\left\langle s_{n}\right\rangle \left[\left(a_{n}-1\right)\ln\rho_{n}-b_{n}\rho_{n}\right]\nonumber \\
 & +\sum_{n=1}^{N}\left\langle 1-s_{n}\right\rangle \left[\left(\overline{a}_{n}-1\right)\ln\rho_{n}-\overline{b}_{n}\rho_{n}\right]\nonumber \\
\propto & \sum_{n=1}^{N}\left[\left\langle s_{n}\right\rangle a_{n}+\left\langle 1-s_{n}\right\rangle \overline{a}_{n}\right]\ln\rho_{n}\nonumber \\
 & -\sum_{n=1}^{N}\left[\left\langle s_{n}\right\rangle b_{n}+\left\langle 1-s_{n}\right\rangle \overline{b}_{n}+\left\langle x_{n}^{2}\right\rangle \right]\rho_{n}\nonumber \\
\propto & \sum_{n=1}^{N}\left(\widetilde{a}_{n}-1\right)\ln\rho_{n}-\sum_{n=1}^{N}\widetilde{b}_{n}\rho_{n}.\label{eq:lnq(rou_n)}
\end{align}
And thus, $\boldsymbol{\rho}$ has a form of product of Gamma distributions
\begin{equation}
q\left(\boldsymbol{\rho}\right)=\prod_{n=1}^{N}\textrm{Ga}\left(\rho_{n};\widetilde{a}_{n},\widetilde{b}_{n}\right),
\end{equation}
where the parameters $\widetilde{a}_{n}$ and $\widetilde{b}_{n}$
are respectively given by
\begin{equation}
\begin{aligned}\widetilde{a}_{n} & =\left\langle s_{n}\right\rangle a_{n}+\left\langle 1-s_{n}\right\rangle \overline{a}_{n}+1,\\
\widetilde{b}_{n} & =\left\langle s_{n}\right\rangle b_{n}+\left\langle 1-s_{n}\right\rangle \overline{b}_{n}+\left\langle x_{n}^{2}\right\rangle .
\end{aligned}
\label{eq:rou_post}
\end{equation}
}

\subsubsection{\textcolor{blue}{Update of $q\left(\boldsymbol{s}\right)$}}

\textcolor{blue}{The posterior distribution $q\left(\boldsymbol{s}\right)$
can be calculated by
\begin{align}
\ln q\left(\boldsymbol{s}\right)\propto & \left\langle \ln p\left(\boldsymbol{v},\boldsymbol{y}\right)\right\rangle _{q\left(\boldsymbol{x}\right)q\left(\boldsymbol{\rho}\right)q\left(\gamma\right)q\left(\boldsymbol{\theta}\right)}\nonumber \\
\propto & \left\langle \ln p\left(\boldsymbol{\rho}\mid\boldsymbol{s}\right)\right\rangle _{q\left(\boldsymbol{\rho}\right)}+\ln p\left(\boldsymbol{s}\right)\nonumber \\
\propto & \sum_{n=1}^{N}s_{n}\ln C_{n}+\sum_{n=1}^{N}\left(1-s_{n}\right)\ln\overline{C}_{n}\nonumber \\
 & +\sum_{n=1}^{N}s_{n}\ln\lambda_{n}+\sum_{n=1}^{N}\left(1-s_{n}\right)\ln\left(1-\lambda_{n}\right)\nonumber \\
\propto & \sum_{n=1}^{N}s_{n}\ln\lambda_{n}C_{n}+\sum_{n=1}^{N}\left(1-s_{n}\right)\ln\left(1-\lambda_{n}\right)\overline{C}_{n}\nonumber \\
\propto & \sum_{n=1}^{N}s_{n}\ln\widetilde{\lambda}_{n}+\sum_{n=1}^{N}\left(1-s_{n}\right)\ln\left(1-\widetilde{\lambda}_{n}\right),\label{eq:lnq(s)}
\end{align}
with $C_{n}=\dfrac{b_{n}^{a_{n}}}{\Gamma\left(a_{n}\right)}\exp\left(\left(a_{n}-1\right)\left\langle \ln\rho_{n}\right\rangle -b_{n}\left\langle \rho_{n}\right\rangle \right)$
and $\overline{C}_{n}=\dfrac{\overline{b}_{n}^{\overline{a}_{n}}}{\Gamma\left(\overline{a}_{n}\right)}\exp\left(\left(\overline{a}_{n}-1\right)\left\langle \ln\rho_{n}\right\rangle -\overline{b}_{n}\left\langle \rho_{n}\right\rangle \right)$.
Here, $\Gamma\left(\cdot\right)$ denotes the gamma function. Hence,
$\boldsymbol{s}$ has a form of product of Bernoulli distributions
\begin{equation}
q\left(\boldsymbol{s}\right)=\prod_{n=1}^{N}\left(\widetilde{\lambda}_{n}\right)^{s_{n}}\left(1-\widetilde{\lambda}_{n}\right)^{1-s_{n}},\label{eq:q(s)}
\end{equation}
where $\widetilde{\lambda}_{n}$ is given by
\begin{equation}
\widetilde{\lambda}_{n}=\frac{\lambda_{n}C_{n}}{\lambda_{n}C_{n}+\left(1-\lambda_{n}\right)\overline{C}_{n}}.\label{eq:lambda_n_tilde}
\end{equation}
}

\subsubsection{\textcolor{blue}{Update of $q\left(\gamma\right)$}}

\textcolor{blue}{The posterior distribution $q\left(\gamma\right)$
is given by
\begin{align}
\ln q\left(\gamma\right)\propto & \left\langle \ln p\left(\boldsymbol{v},\boldsymbol{y}\right)\right\rangle _{q\left(\boldsymbol{x}\right)q\left(\boldsymbol{\rho}\right)q\left(\boldsymbol{s}\right)q\left(\boldsymbol{\theta}\right)}\nonumber \\
\propto & \left\langle \ln p\left(\boldsymbol{y}\mid\boldsymbol{x},\boldsymbol{\theta},\gamma\right)\right\rangle _{q\left(\boldsymbol{x}\right)q\left(\boldsymbol{\theta}\right)}+\ln p\left(\gamma\right)\nonumber \\
\propto & M\ln\gamma-\gamma\left\langle \left\Vert \boldsymbol{y}-\mathbf{F}\left(\boldsymbol{\theta}\right)\boldsymbol{x}\right\Vert ^{2}\right\rangle _{q\left(\boldsymbol{x}\right)q\left(\boldsymbol{\theta}\right)}\nonumber \\
 & +\left(c-1\right)\ln\gamma-d\gamma\nonumber \\
\propto & \left(M+c-1\right)\ln\gamma\nonumber \\
 & -\left(d+\left\langle \left\Vert \boldsymbol{y}-\mathbf{F}\left(\boldsymbol{\theta}\right)\boldsymbol{x}\right\Vert ^{2}\right\rangle _{q\left(\boldsymbol{x}\right)q\left(\boldsymbol{\theta}\right)}\right)\gamma\nonumber \\
\propto & \left(\widetilde{c}-1\right)\ln\gamma-\widetilde{d}\gamma.\label{eq:lnq(gamma)}
\end{align}
Thus, $\gamma$ follows a Gamma distribution
\begin{equation}
q\left(\gamma\right)=\textrm{Ga}\left(\gamma;\widetilde{c},\widetilde{d}\right),\label{eq:q(gamma)}
\end{equation}
where the parameters $\widetilde{c}$ and $\widetilde{d}$ are respectively
given by
\begin{align}
\widetilde{c} & =c+M,\label{eq:gamma_post_old}\\
\widetilde{d} & =d+\left\langle \left\Vert \boldsymbol{y}-\mathbf{F}\left(\boldsymbol{\theta}\right)\boldsymbol{x}\right\Vert ^{2}\right\rangle _{q\left(\boldsymbol{x}\right)q\left(\boldsymbol{\theta}\right)}.\nonumber 
\end{align}
}

\subsubsection{\textcolor{blue}{Update of $q\left(\boldsymbol{\theta}\right)$}}

\textcolor{blue}{The posterior distribution $q\left(\boldsymbol{\theta}^{j}\right)$
can be derived as
\begin{align}
\ln q\left(\boldsymbol{\theta}^{j}\right)\propto & \left\langle \ln p\left(\boldsymbol{v},\boldsymbol{y}\right)\right\rangle _{q\left(\boldsymbol{x}\right)q\left(\boldsymbol{\rho}\right)q\left(\boldsymbol{s}\right)q\left(\gamma\right)\Pi_{i\neq j}q\left(\boldsymbol{\theta}^{i}\right)}\nonumber \\
\propto & \left\langle \ln p\left(\boldsymbol{y}\mid\boldsymbol{x},\boldsymbol{\theta},\gamma\right)\right\rangle _{q\left(\boldsymbol{x}\right)q\left(\gamma\right)\Pi_{i\neq j}q\left(\boldsymbol{\theta}^{i}\right)}\nonumber \\
 & +\ln p\left(\boldsymbol{\theta}^{j}\right),\forall j\in\left\{ 1,\ldots,B\right\} .\label{eq:q(theta)}
\end{align}
}Here we assume that the prior distribution of $\boldsymbol{\theta}^{j}$
is $p\left(\boldsymbol{\theta}^{j}\right)=\mathcal{N}\bigl(\boldsymbol{\theta}^{j};\overline{\boldsymbol{\theta}}^{j},1/\kappa^{j}\mathbf{I}_{N}\bigr)$,
where $\overline{\boldsymbol{\theta}}^{j}$ is the initial value of
the dynamic grid and $\kappa^{j}$ is the precision of $\boldsymbol{\theta}^{j}$.
Note that it is natural to assume a Gaussian prior for $\boldsymbol{\theta}^{j}$
when the initial value $\overline{\boldsymbol{\theta}}^{j}$ is a
uniform sampling grid and the precision $\kappa^{j}$ is sufficiently
small.

The expectations used in the above update expressions are summarized
as follows:\textcolor{blue}{
\[
\begin{aligned}\left\langle \rho_{n}\right\rangle = & \begin{aligned}\dfrac{\widetilde{a}_{n}}{\widetilde{b}_{n}} &  & \left\langle \boldsymbol{\rho}\right\rangle = & \left[\bigl\langle\rho_{1}\bigr\rangle,\ldots,\bigl\langle\rho_{N}\bigr\rangle\right]^{T} & \left\langle s_{n}\right\rangle  & =\widetilde{\lambda}_{n}\end{aligned}
\\
\left\langle \gamma\right\rangle = & \begin{aligned}\dfrac{\widetilde{c}}{\widetilde{d}} &  & \left\langle x_{n}^{2}\right\rangle = & \left|\mu_{x,n}\right|^{2}+\Sigma_{x,n,n} & \left\langle \ln\rho_{n}\right\rangle = & \psi\left(\widetilde{a}_{n}\right)-\ln\widetilde{b}_{n},\end{aligned}
\end{aligned}
\]
where $\mu_{x,n}$ is the $n\textrm{-th}$ element of $\boldsymbol{\mu}_{x}$,
$\boldsymbol{\Sigma}_{x,n,n}$ is the $n\textrm{-th}$ diagonal element
of $\boldsymbol{\Sigma}_{x}$, and $\psi\left(\cdot\right)\triangleq d\ln\left(\Gamma\left(\cdot\right)\right)$
denotes the logarithmic derivative of the gamma function.}

However, since $\mathbf{F}\left(\boldsymbol{\theta}\right)$ is nonlinear
w.r.t $\boldsymbol{\theta}$, we cannot obtain the closed-from expressions
of $q\left(\boldsymbol{x}\right)$, $q\left(\gamma\right)$, and $q\left(\boldsymbol{\theta}\right)$.
Although some particle-based methods \cite{Danescu_particle1,Zhou_particle3}
were proposed to address this problem, the time complexity of these
methods is usually very high due to the large number of random sampling.
In the next subsection, we will elaborate on how to compute the posteriors
approximately based on the successive linear approximation approach.

\subsection{Successive Linear Approximation}

\begin{algorithm}[t]
\begin{singlespace}
{\small{}\caption{\label{LAVBI}SLA-VBI algorithm}
}{\small\par}

\textbf{Input:} $\boldsymbol{y}$, initial grid $\overline{\boldsymbol{\theta}}$,
iteration number $I=I_{1}+I_{2}$.

\textbf{Output:} $\boldsymbol{x}^{*}$, $\boldsymbol{s}^{*}$, and
$\boldsymbol{\theta}^{*}$.

\begin{algorithmic}[1]

\STATE \textbf{\% Stage 1: Initialization}

\STATE Initialize $\hat{\boldsymbol{\mu}}_{\theta^{j}}=\overline{\boldsymbol{\theta}}^{j},\forall j$
and fix $\boldsymbol{\theta=\overline{\boldsymbol{\theta}}}$.

\FOR{${\color{blue}{\color{black}i=1,\cdots,I_{1}}}$}

\STATE Optimize $q\left(\boldsymbol{x}\right)$, $q\left(\boldsymbol{\rho}\right)$,
$q\left(\boldsymbol{s}\right)$, $q\left(\gamma\right)$, using (\ref{eq:x_post}),
(\ref{eq:rou_post}), (\ref{eq:lambda_n_tilde}), (\ref{eq:gamma_post}).

\ENDFOR

\STATE \textbf{\% Stage 2: Successive Linear Approximation VBI}

\FOR{${\color{blue}{\color{black}i=I_{1}+1,\cdots,I}}$}

\STATE Linear approximation using (\ref{eq:linearization_vector})
and (\ref{eq:linearization_matrix}).

\STATE Optimize $q\left(\boldsymbol{x}\right)$, $q\left(\boldsymbol{\rho}\right)$,
$q\left(\boldsymbol{s}\right)$, $q\left(\gamma\right)$, $q\left(\boldsymbol{\theta}\right)$
in an alternating fashion, using (\ref{eq:x_post}), (\ref{eq:rou_post}),
(\ref{eq:lambda_n_tilde}), (\ref{eq:gamma_post}), (\ref{eq:theta_post}).

\STATE Let $\hat{\boldsymbol{\mu}}_{\theta^{j}}=\boldsymbol{\mu}_{\theta^{j}},\forall j$,
using (\ref{eq:theta_post}).

\ENDFOR

\STATE Output $\boldsymbol{x}^{*}=\boldsymbol{\mu}_{x}$, $s_{n}^{*}=\widetilde{\lambda}_{n},\forall n$,
and $\mathbf{\boldsymbol{\theta}^{\mathit{j}}}^{*}=\boldsymbol{\mu}_{\theta^{j}},\forall j$.

\end{algorithmic}
\end{singlespace}
\end{algorithm}
In order to perform Bayesian inference in closed form, one common
solution is to approximate the nonlinear mapping $\boldsymbol{\theta}\rightarrow\mathbf{F}\left(\boldsymbol{\theta}\right)$
to a linear mapping. \textcolor{blue}{To simplify the notation}, $\hat{\mu}_{\theta_{n}^{j}}$,
$\hat{\boldsymbol{\mu}}_{\theta^{j}}$, $\hat{\boldsymbol{\mu}}_{\theta_{n}}$,
and $\hat{\boldsymbol{\mu}}_{\theta}$ are used to denote the posterior
means of $\theta_{n}^{j}$, $\boldsymbol{\theta}^{j}$, $\boldsymbol{\theta}_{n}$,
and $\boldsymbol{\theta}$, respectively, which are obtained in the
latest iteration. Using linearization, the basis vector $\boldsymbol{\varPhi}\left(\boldsymbol{\theta}_{n}\right)$
is approximated to
\begin{equation}
\boldsymbol{\varPhi}\left(\boldsymbol{\theta}_{n}\right)\approx\boldsymbol{\varPhi}\left(\hat{\boldsymbol{\mu}}_{\theta_{n}}\right)+\sum_{j=1}^{B}\frac{\partial\boldsymbol{\varPhi}\left(\hat{\boldsymbol{\mu}}_{\theta_{n}}\right)}{\partial\theta_{n}^{j}}\left(\theta_{n}^{j}-\hat{\mu}_{\theta_{n}^{j}}\right),\forall n.\label{eq:linearization_vector}
\end{equation}
Then the measurement matrix $\mathbf{F}\left(\boldsymbol{\theta}\right)$
is approximated to 
\begin{equation}
\mathbf{F}\left(\boldsymbol{\theta}\right)\approx\mathbf{F}\left(\hat{\boldsymbol{\mu}}_{\theta}\right)+\sum_{j=1}^{B}\mathbf{A^{\mathit{j}}}\textrm{diag}\left(\boldsymbol{\theta}^{j}-\hat{\boldsymbol{\mu}}_{\theta^{j}}\right)\triangleq\overline{\mathbf{F}}\left(\boldsymbol{\theta}\right),\label{eq:linearization_matrix}
\end{equation}
where $\mathbf{A^{\mathit{j}}}\triangleq\left[\frac{\partial\boldsymbol{\varPhi}\left(\hat{\boldsymbol{\mu}}_{\theta_{1}}\right)}{\partial\theta_{1}^{j}},\ldots,\frac{\partial\boldsymbol{\varPhi}\left(\hat{\boldsymbol{\mu}}_{\theta_{N}}\right)}{\partial\theta_{N}^{j}}\right],j\in\left\{ 1,\ldots,B\right\} $.
\textcolor{blue}{We can further obtain the statistical property of
$\overline{\mathbf{F}}\left(\boldsymbol{\theta}\right)$ as
\begin{align}
\left\langle \overline{\mathbf{F}}\left(\boldsymbol{\theta}\right)\right\rangle _{q\left(\boldsymbol{\theta}\right)}= & \mathbf{F}\left(\hat{\boldsymbol{\mu}}_{\theta}\right),\nonumber \\
\left\langle \overline{\mathbf{F}}\left(\boldsymbol{\theta}\right)^{H}\overline{\mathbf{F}}\left(\boldsymbol{\theta}\right)\right\rangle _{q\left(\boldsymbol{\theta}\right)}= & \mathbf{F}\left(\hat{\boldsymbol{\mu}}_{\theta}\right)^{H}\mathbf{F}\left(\hat{\boldsymbol{\mu}}_{\theta}\right)\nonumber \\
 & +\sum_{j=1}^{B}\left(\mathbf{A^{\mathit{j}}}^{H}\mathbf{A^{\mathit{j}}}\right)\odot\boldsymbol{\Sigma}_{\theta^{j}}\triangleq\mathbf{H}_{x},\label{eq:statis_Ftheta}
\end{align}
where $\boldsymbol{\Sigma}_{\theta^{j}}$ is the posterior covariance
matrix of $\boldsymbol{\theta}^{j}$ obtained in (\ref{eq:theta_post}).
Based on the linear approximation of $\mathbf{F}\left(\boldsymbol{\theta}\right)$
and the statistical property of $\overline{\mathbf{F}}\left(\boldsymbol{\theta}\right)$,
the variational distributions $q\left(\boldsymbol{x}\right)$, $q\left(\gamma\right)$,
and $q\left(\boldsymbol{\theta}\right)$ can be derived in closed
form.}

\textcolor{blue}{Substituting (\ref{eq:linearization_matrix}) into
(\ref{eq:x_post_old}) and using (\ref{eq:statis_Ftheta}), the approximate
posterior mean and covariance matrix of $\boldsymbol{x}$ can be rewritten
into
\begin{equation}
\begin{aligned}\boldsymbol{\mu}_{x}= & \boldsymbol{\Sigma}_{x}\left\langle \gamma\right\rangle \mathbf{F}\left(\hat{\boldsymbol{\mu}}_{\theta}\right)^{H}\boldsymbol{y},\\
\boldsymbol{\Sigma}_{x}= & \left(\left\langle \gamma\right\rangle \mathbf{H}_{x}+\textrm{diag}\left(\bigl\langle\boldsymbol{\rho}\bigr\rangle\right)\right)^{-1}.
\end{aligned}
\label{eq:x_post}
\end{equation}
Combining (\ref{eq:linearization_matrix}) with (\ref{eq:gamma_post_old}),
the hyper-parameters of $q\left(\gamma\right)$ are updated as follows:
\begin{align}
\widetilde{c}= & c+M,\nonumber \\
\widetilde{d}= & d+\boldsymbol{y}^{H}\boldsymbol{y}-\boldsymbol{\mu}_{x}^{H}\mathbf{F}\left(\hat{\boldsymbol{\mu}}_{\theta}\right)^{H}\boldsymbol{y}-\boldsymbol{y}^{H}\mathbf{F}\left(\hat{\boldsymbol{\mu}}_{\theta}\right)\boldsymbol{\mu}_{x}\nonumber \\
 & +\boldsymbol{\mu}_{x}^{H}\mathbf{H}_{x}\boldsymbol{\mu}_{x}+\textrm{Tr}\Bigl(\mathbf{H}_{x}\boldsymbol{\Sigma}_{x}\Bigr).\label{eq:gamma_post}
\end{align}
Substituting (\ref{eq:linearization_matrix}) into (\ref{eq:q(theta)}),
$q\left(\boldsymbol{\theta}^{j}\right)$ can be derived as
\begin{align}
\ln q\left(\boldsymbol{\theta}^{j}\right)\propto & -\left\langle \gamma\right\rangle \left\langle \left\Vert \boldsymbol{y}-\overline{\mathbf{F}}\left(\boldsymbol{\theta}\right)\boldsymbol{x}\right\Vert ^{2}\right\rangle _{q\left(\boldsymbol{x}\right)\Pi_{i\neq j}q\left(\boldsymbol{\theta}^{i}\right)}\nonumber \\
 & +\ln\mathcal{N}\left(\boldsymbol{\theta}^{j};\overline{\boldsymbol{\theta}}^{j},1/\kappa^{j}\mathbf{I}_{N}\right)\nonumber \\
\propto & -\frac{1}{2}\left(\boldsymbol{\theta}^{j}-\hat{\boldsymbol{\mu}}_{\theta^{j}}\right)^{T}\left\langle \gamma\right\rangle \mathbf{H}_{\theta^{j}}\left(\boldsymbol{\theta}^{j}-\hat{\boldsymbol{\mu}}_{\theta^{j}}\right)\nonumber \\
 & +\left(\boldsymbol{\theta}^{j}-\hat{\boldsymbol{\mu}}_{\theta^{j}}\right)^{T}\left\langle \gamma\right\rangle \boldsymbol{g}_{\theta^{j}}\nonumber \\
 & +\ln\mathcal{N}\left(\boldsymbol{\theta}^{j};\overline{\boldsymbol{\theta}}^{j},1/\kappa^{j}\mathbf{I}_{N}\right)\nonumber \\
\propto & \ln\mathcal{N}\left(\boldsymbol{\theta}^{j};\mathbf{H}_{\theta^{j}}^{-1}\boldsymbol{g}_{\theta^{j}}+\hat{\boldsymbol{\mu}}_{\theta^{j}},\left\langle \gamma\right\rangle ^{-1}\mathbf{H}_{\theta^{j}}^{-1}\right)\nonumber \\
 & +\ln\mathcal{N}\left(\boldsymbol{\theta}^{j};\overline{\boldsymbol{\theta}}^{j},1/\kappa^{j}\mathbf{I}_{N}\right)\nonumber \\
\propto & \ln\mathcal{N}\left(\boldsymbol{\theta}^{j};\boldsymbol{\mu}_{\theta^{j}},\boldsymbol{\Sigma}_{\theta^{j}}\right),\forall j,\label{eq:lnq(theta_j_final)}
\end{align}
}where the immediate variables has been defined to simplify notations:
\begin{equation}
\begin{aligned}\mathbf{H}_{\theta^{j}}= & 2\mathfrak{Re}\left\{ \left(\boldsymbol{\mu}_{x}\boldsymbol{\mu}_{x}^{H}+\boldsymbol{\Sigma}_{x}\right)^{T}\odot\left(\mathbf{A^{\mathit{j}}}^{H}\mathbf{A^{\mathit{j}}}\right)\right\} ,\forall j,\\
\boldsymbol{g}_{\theta^{j}}= & 2\mathfrak{Re}\left\{ \textrm{diag}\left(\boldsymbol{\mu}_{x}\right)^{H}\mathbf{A^{\mathit{j}}}^{H}\left(\boldsymbol{y}-\mathbf{F}\left(\hat{\boldsymbol{\mu}}_{\theta}\right)\boldsymbol{\mu}_{x}\right)\right\} \\
 & -2\mathfrak{Re}\left\{ \textrm{diag}\left(\mathbf{A^{\mathit{j}}}^{H}\mathbf{F}\left(\hat{\boldsymbol{\mu}}_{\theta}\right)\boldsymbol{\Sigma}_{x}\right)\right\} ,\forall j,
\end{aligned}
\end{equation}
and the approximate posterior mean and covariance matrix of $\boldsymbol{\theta}^{j}$
are respectively given by
\begin{equation}
\begin{aligned}\boldsymbol{\mu}_{\theta^{j}} & =\boldsymbol{\Sigma}_{\theta^{j}}\left(\left\langle \gamma\right\rangle \left(\boldsymbol{g}_{\theta^{j}}+\mathbf{H}_{\theta^{j}}\hat{\boldsymbol{\mu}}_{\theta^{j}}\right)+\kappa^{j}\overline{\boldsymbol{\theta}}^{j}\right),\forall j,\\
\boldsymbol{\Sigma}_{\theta^{j}} & =\left(\left\langle \gamma\right\rangle \mathbf{H}_{\theta^{j}}+\kappa^{j}\mathbf{I}_{N}\right)^{-1},\forall j.
\end{aligned}
\label{eq:theta_post}
\end{equation}
\textcolor{blue}{Please refer to Appendix \ref{subsec:Derivation-of-equations}
for more details of the derivation of (\ref{eq:lnq(theta_j_final)})
- (\ref{eq:theta_post}).}

The complete algorithm is shown in Algorithm \ref{LAVBI}, which contains
two main stages. Specifically, in stage 1, we keep the grid parameters
$\boldsymbol{\theta}$ fixed and optimize other variational distributions
to find a good initial value for other variables. In stage 2, we update
the approximate posterior distribution of all variables based on the
successive linear approximation approach. Note that the linear approximation
approach used in our proposed algorithm is quite different from that
in the OGSBI algorithm. The OGSBI only uses the first order approximation
of the true observation model at the initial sampling grid. In contrast,
our proposed algorithm uses the linear approximation based on the
latest updated grid, and thus the approximate error in (\ref{eq:linearization_vector})
will decease gradually during iterations. Furthermore, the error distribution
of grid parameters is considered during the algorithm design, and
thus the proposed algorithm can provide a Bayesian estimation of grid
parameters. However, the OGSBI and other EM-based methods can only
give a point estimation of grid parameters. Therefore, our proposed
algorithm can achieve a better performance than these methods.

The complexity of the proposed algorithm is dominated by the matrix
inverse operations in (\ref{eq:x_post}) and (\ref{eq:theta_post}),
which is $\Theta\left(N^{3}\right)$. However, when $N$ is large,
it is very time-consuming to obtain the inverse of large scale matrices.
In the next subsection, we will propose an inverse-free algorithm
with lower complexity based on the MM framework.

\subsection{IFSLA-VBI Algorithm}

Recalling (\ref{eq:x_post}) and (\ref{eq:theta_post}), we find that
$\boldsymbol{\mu}_{x}$ and $\boldsymbol{\mu}_{\theta^{j}}$ are the
global optimal solutions of the following minimization problems:
\begin{align}
\boldsymbol{\mu}_{x} & =\boldsymbol{\Sigma}_{x}\left\langle \gamma\right\rangle \mathbf{F}\left(\hat{\boldsymbol{\mu}}_{\theta}\right)^{H}\boldsymbol{y}\nonumber \\
 & =\min_{\boldsymbol{\mu}_{x}}\left(\boldsymbol{\mu}_{x}^{H}\mathbf{W}_{x}\boldsymbol{\mu}_{x}-2\mathfrak{Re}\left\{ \boldsymbol{\mu}_{x}^{H}\boldsymbol{b}_{x}\right\} \right)\triangleq\min_{\boldsymbol{\mu}_{x}}\varphi\left(\boldsymbol{\mu}_{x}\right),\nonumber \\
\boldsymbol{\mu}_{\theta^{j}} & =\boldsymbol{\Sigma}_{\theta^{j}}\left(\left\langle \gamma\right\rangle \left(\boldsymbol{g}_{\theta^{j}}+\mathbf{H}_{\theta^{j}}\hat{\boldsymbol{\mu}}_{\theta^{j}}\right)+\kappa^{j}\overline{\boldsymbol{\theta}}^{j}\right)\nonumber \\
 & =\min_{\boldsymbol{\mu}_{\theta^{j}}}\left(\boldsymbol{\mu}_{\theta^{j}}^{T}\mathbf{W}_{\theta^{j}}\boldsymbol{\mu}_{\theta^{j}}-2\boldsymbol{\mu}_{\theta^{j}}^{T}\boldsymbol{b}_{\theta^{j}}\right)\triangleq\min_{\boldsymbol{\mu}_{\theta^{j}}}\psi\left(\boldsymbol{\mu}_{\theta^{j}}\right),\label{eq:minimun_problem}
\end{align}
where $\mathbf{W}_{x}\triangleq\boldsymbol{\Sigma}_{x}^{-1}=\left\langle \gamma\right\rangle \mathbf{H}_{x}+\textrm{diag}\left(\left\langle \boldsymbol{\rho}\right\rangle \right)$,
$\mathbf{W}_{\theta^{j}}\triangleq\boldsymbol{\Sigma}_{\theta^{j}}^{-1}=\left\langle \gamma\right\rangle \mathbf{H}_{\theta^{j}}+\kappa^{j}\mathbf{I}_{N}$,
$\boldsymbol{b}_{x}=\left\langle \gamma\right\rangle \mathbf{F}\left(\hat{\boldsymbol{\mu}}_{\theta}\right)^{H}\boldsymbol{y}$,
and $\boldsymbol{b}_{\theta^{j}}=\left\langle \gamma\right\rangle \left(\boldsymbol{g}_{\theta^{j}}+\mathbf{H}_{\theta^{j}}\hat{\boldsymbol{\mu}}_{\theta^{j}}\right)+\kappa^{j}\overline{\boldsymbol{\theta}}^{j}$.

The objective functions $\varphi\left(\boldsymbol{\mu}_{x}\right)$
and $\psi\left(\boldsymbol{\mu}_{\theta^{j}}\right)$ are convex with
bounded curvature. And thus, it is suitable to employ the MM framework
to find the global optimal solutions without matrix inverse operation.
Specifically, the surrogate functions for $\varphi\left(\boldsymbol{\mu}_{x}\right)$
and $\psi\left(\boldsymbol{\mu}_{\theta^{j}}\right)$ can be constructed
by resorting to the following lemma \cite{Sun_MM,Duan_IFSBL}:
\begin{lem}
For any continuously differentiable function $f:\mathbb{C}^{N}\rightarrow\mathbb{C}$
with a continuous gradient, we have
\begin{equation}
f\left(\boldsymbol{u}\right)\leq f\left(\boldsymbol{v}\right)+\left(\boldsymbol{u}-\boldsymbol{v}\right)^{H}\nabla f\left(\boldsymbol{v}\right)+\left(\boldsymbol{u}-\boldsymbol{v}\right)^{H}\mathbf{T}\left(\boldsymbol{u}-\boldsymbol{v}\right),\label{eq:L-smooth}
\end{equation}
for any $\boldsymbol{u},\boldsymbol{v}\in\mathbb{C}^{N}$ and $\mathbf{T}\succcurlyeq\frac{\nabla^{2}f\left(\boldsymbol{x}\right)}{2},\forall\boldsymbol{x}$.
\end{lem}
In the majorization step, we construct the surrogate function by applying
(\ref{eq:L-smooth}). Specifically, in the $t\textrm{-th}$ MM iteration,
the surrogate functions for $\varphi\left(\boldsymbol{\mu}_{x}\right)$
and $\psi\left(\boldsymbol{\mu}_{\theta^{j}}\right)$ are respectively
given by
\begin{align}
\mathring{\varphi}\left(\boldsymbol{\mu}_{x};\boldsymbol{\mu}_{x}^{\left(t\right)}\right)= & \varphi\left(\boldsymbol{\mu}_{x}^{\left(t\right)}\right)\nonumber \\
+ & 2\mathfrak{Re}\left\{ \left(\boldsymbol{\mu}_{x}-\boldsymbol{\mu}_{x}^{\left(t\right)}\right)^{H}\left(\mathbf{W}_{x}\boldsymbol{\mu}_{x}^{\left(t\right)}-\boldsymbol{b}_{x}\right)\right\} \nonumber \\
+ & \left\langle \gamma\right\rangle L_{x}^{\left(t\right)}\left\Vert \boldsymbol{\mu}_{x}-\boldsymbol{\mu}_{x}^{\left(t\right)}\right\Vert ^{2}\nonumber \\
+ & \left(\boldsymbol{\mu}_{x}-\boldsymbol{\mu}_{x}^{\left(t\right)}\right)^{H}\textrm{diag}\left(\left\langle \boldsymbol{\rho}\right\rangle \right)\left(\boldsymbol{\mu}_{x}-\boldsymbol{\mu}_{x}^{\left(t\right)}\right),\nonumber \\
\mathring{\psi}\left(\boldsymbol{\mu}_{\theta^{j}};\boldsymbol{\mu}_{\theta^{j}}^{\left(t\right)}\right)= & \psi\left(\boldsymbol{\mu}_{\theta^{j}}^{\left(t\right)}\right)\nonumber \\
+ & 2\left(\boldsymbol{\mu}_{\theta^{j}}-\boldsymbol{\mu}_{\theta^{j}}^{\left(t\right)}\right)^{T}\left(\mathbf{W}_{\theta^{j}}\boldsymbol{\mu}_{\theta^{j}}^{\left(t\right)}-\boldsymbol{b}_{\theta^{j}}\right)\nonumber \\
+ & \left(\left\langle \gamma\right\rangle L_{\theta^{j}}^{\left(t\right)}+\kappa^{j}\right)\left\Vert \boldsymbol{\mu}_{\theta^{j}}-\boldsymbol{\mu}_{\theta^{j}}^{\left(t\right)}\right\Vert ^{2}.\label{eq:surrogate functions}
\end{align}
where $L_{x}^{\left(t\right)}$ and $L_{\theta^{j}}^{\left(t\right)}$
need to satisfy
\begin{equation}
\begin{aligned}L_{x}^{\left(t\right)}\mathbf{I}_{N} & \succcurlyeq\mathbf{H}_{x},\\
L_{\theta^{j}}^{\left(t\right)}\mathbf{I}_{N} & \succcurlyeq\mathbf{H}_{\theta^{j}}.
\end{aligned}
\label{eq:inequality}
\end{equation}
In the minimization step, we minimize the surrogate functions, which
leads to the following update:
\begin{equation}
\begin{aligned}\boldsymbol{\mu}_{x}^{\left(t+1\right)} & =\mathbf{\Lambda}_{x}^{\left(t\right)}\boldsymbol{\zeta}_{x}^{\left(t\right)},\\
\boldsymbol{\mu}_{\theta^{j}}^{\left(t+1\right)} & =\Lambda_{\theta^{j}}^{\left(t\right)}\boldsymbol{\zeta}_{\theta^{j}}^{\left(t\right)},
\end{aligned}
\label{eq:update rule}
\end{equation}
where the immediate variables has been defined to simplify notations:
\begin{align}
\boldsymbol{\zeta}_{x}^{\left(t\right)} & =\left(\left\langle \gamma\right\rangle L_{x}^{(t)}\mathbf{I}_{N}+\textrm{diag}\left(\left\langle \boldsymbol{\rho}\right\rangle \right)-\mathbf{W}_{x}\right)\boldsymbol{\mu}_{x}^{\left(t\right)}+\boldsymbol{b}_{x},\nonumber \\
\mathbf{\Lambda}_{x}^{\left(t\right)} & =\left(\left\langle \gamma\right\rangle L_{x}^{(t)}\mathbf{I}_{N}+\textrm{diag}\left(\left\langle \boldsymbol{\rho}\right\rangle \right)\right)^{-1},\nonumber \\
\boldsymbol{\zeta}_{\theta^{j}}^{\left(t\right)} & =\left(\left(\left\langle \gamma\right\rangle L_{\theta^{j}}^{(t)}+\kappa^{j}\right)\mathbf{I}_{N}-\mathbf{W}_{\theta^{j}}\right)\boldsymbol{\mu}_{\theta^{j}}^{\left(t\right)}+\boldsymbol{b}_{\theta^{j}},\nonumber \\
\Lambda_{\theta^{j}}^{\left(t\right)} & =\left(\left\langle \gamma\right\rangle L_{\theta^{j}}^{(t)}+\kappa^{j}\right)^{-1}.\label{eq:immediate_variables}
\end{align}
We set $L_{x}^{\left(t\right)}=L_{x}^{0}\left(1+c_{x}\right)^{t}$
and $L_{\theta^{j}}^{\left(t\right)}=L_{\theta^{j}}^{0}\left(1+c_{\theta^{j}}\right)^{t}$,
where $L_{x}^{0},L_{\theta^{j}}^{0},c_{x},c_{\theta^{j}}>0$ are hyper-parameters
determined by the structure of the sensing matrix. A good choice for
hyper-parameters is $L_{x}^{0}=\lambda_{\textrm{max}}\bigl(\mathbf{F}\left(\boldsymbol{\theta}\right)^{H}\mathbf{F}\left(\boldsymbol{\theta}\right)\bigr)|{}_{\boldsymbol{\theta}=\boldsymbol{\overline{\theta}}}$,
$L_{\theta^{j}}^{0}=\lambda_{\textrm{max}}\left(\mathbf{H}_{\theta^{j}}\right)|{}_{\boldsymbol{\theta}^{j}=\boldsymbol{\overline{\theta}}^{j}}$,
and $c_{x},c_{\theta^{j}}\in\left(0,0.1\right)$, where $\lambda_{\textrm{max}}\left(\cdot\right)$
denotes the largest eigenvalue of the given matrix. In the initialization
stage, $L_{x}^{0}$ and $L_{\theta^{j}}^{0}$ can be calculated by
singular value decomposition (SVD) based on the initial grid $\boldsymbol{\overline{\theta}}$.
It is also possible to use the data-driven approach to learning these
hyper-parameters for even better performance when the training data
is available. Note that the inequality in (\ref{eq:inequality}) holds
strictly when $t$ is sufficiently large. Therefore, the update rule
in (\ref{eq:update rule}) guarantees the convergence of the algorithm
to the global optimal of (\ref{eq:minimun_problem}) when $t$ is
sufficiently large.

As $\mathbf{\Lambda}_{x}^{\left(t\right)}$ is a diagonal matrix,
and $\boldsymbol{\zeta}_{x}^{\left(t\right)}$ and $\boldsymbol{\zeta}_{\theta^{j}}^{\left(t\right)}$
are computed by the matrix-vector multiplications, the computational
complexity of the matrix inverse is reduced to $\Theta\left(N^{2}T\right)$,
where $T$ represents the number of local iterations in (\ref{eq:update rule}). 

Moreover, the posterior covariance matrices $\boldsymbol{\Sigma}_{x}$
and $\boldsymbol{\Sigma}_{\theta^{j}}$ are approximated by setting
non-diagonal elements to be zero, i.e.,
\begin{equation}
\begin{aligned}\boldsymbol{\Sigma}_{x} & =\textrm{diag}\left(\left[\mathrm{W}_{x,1,1}^{-1},\ldots,\mathrm{W}_{x,N,N}^{-1}\right]\right),\\
\boldsymbol{\Sigma}_{\theta^{j}} & =\textrm{diag}\left(\left[\mathrm{W}_{\theta^{j},1,1}^{-1},\ldots,\mathrm{W}_{\theta^{j},N,N}^{-1}\right]\right),
\end{aligned}
\end{equation}
where $\mathrm{W}_{x,n,n}$ and $\mathrm{W}_{\theta^{j},n,n}$ are
the $n\textrm{-th}$ diagonal elements of $\mathbf{W}_{x}$ and $\mathbf{W}_{\theta^{j}}$,
respectively. 

\begin{algorithm}[t]
\begin{singlespace}
{\small{}\caption{\label{IF-LAVBI}IFSLA-VBI algorithm}
}{\small\par}

\textbf{Input:} $\boldsymbol{y}$, initial grid $\overline{\boldsymbol{\theta}}$,
iteration number $I=I_{1}+I_{2}$, local iteration number $T$.

\textbf{Output:} $\boldsymbol{x}^{*}$, $\boldsymbol{s}^{*}$, and
$\boldsymbol{\theta}^{*}$.

\begin{algorithmic}[1]

\STATE \textbf{\% Stage 1: Initialization}

\STATE Initialize $\hat{\boldsymbol{\mu}}_{\theta^{j}}=\overline{\boldsymbol{\theta}}^{j},\forall j$,
and fix $\boldsymbol{\theta=\overline{\boldsymbol{\theta}}}$.

\FOR{${\color{blue}{\color{black}i=1,\cdots,I_{1}}}$}

\STATE Optimize $q\left(\boldsymbol{x}\right)$, using the MM framework
to approximate the matrix inverse.

\STATE Optimize $q\left(\boldsymbol{\rho}\right)$, $q\left(\boldsymbol{s}\right)$,
$q\left(\gamma\right)$, using (\ref{eq:rou_post}), (\ref{eq:lambda_n_tilde}),
(\ref{eq:gamma_post}).

\ENDFOR

\STATE \textbf{\% Stage 2: IFSLA-VBI Algorithm}

\FOR{${\color{blue}{\color{black}i=I_{1}+1,\cdots,I}}$}

\STATE Linear approximation using (\ref{eq:linearization_vector})
and (\ref{eq:linearization_matrix}).

\STATE \textbf{\% The MM framework to optimize $q\left(\boldsymbol{x}\right)$
and $q\left(\boldsymbol{\theta}\right)$}

\WHILE{not converge and $t\leq T$}

\STATE Majorization: construct surrogate functions in (\ref{eq:surrogate functions}).

\STATE Minimization: obey the update rule in (\ref{eq:update rule}).

\ENDWHILE

\STATE Optimize $q\left(\boldsymbol{\rho}\right)$, $q\left(\boldsymbol{s}\right)$,
$q\left(\gamma\right)$, using (\ref{eq:rou_post}), (\ref{eq:lambda_n_tilde}),
(\ref{eq:gamma_post}).

\STATE Let $\hat{\boldsymbol{\mu}}_{\theta^{j}}=\boldsymbol{\mu}_{\theta^{j}},\forall j$,
using (\ref{eq:theta_post}).

\ENDFOR

\STATE Output $\boldsymbol{x}^{*}=\boldsymbol{\mu}_{x}$, $s_{n}^{*}=\widetilde{\lambda}_{n},\forall n$,
and $\mathbf{\boldsymbol{\theta}^{\mathit{j}}}^{*}=\boldsymbol{\mu}_{\theta^{j}},\forall j$.

\end{algorithmic}
\end{singlespace}
\end{algorithm}
The simplified inverse-free successive linear approximation VBI algorithm,
hereafter referred to as IFSLA-VBI, is shown in Algorithm \ref{IF-LAVBI}.
The IFSLA-VBI algorithm can achieve a better trade-off between performance
and complexity by controlling the number of local iterations $T$
according to the structure of sensing matrix. Specifically, for the
case of well-conditioned sensing matrices, the IFSLA-VBI algorithm
only requires a small $T$ to reach convergence. In this case, the
computational overhead can be reduced greatly. On the other hand,
for the case of ill-conditioned sensing matrices, a relatively large
$T$ is usually needed.

\subsection{Complexity Comparison}

\textcolor{blue}{We analyze the computational complexity of the proposed
IFSLA-VBI and the state-of-the-art Turbo-CS and Turbo-VBI.} In the
IFSLA-VBI algorithm, the matrix inverse operation has been simplified
into some matrix-vector multiplications, whose complexity is $\Theta\left(N^{2}T\right)$.
The matrix multiplication $\mathbf{F}\left(\hat{\boldsymbol{\mu}}_{\theta}\right)^{H}\mathbf{F}\left(\hat{\boldsymbol{\mu}}_{\theta}\right)$
has a computational complexity scaling as $\Theta\left(N^{2}M\right)$.
Besides, the multiplication of two $N\times N$ matrices, i.e., $\bigl\{\mathbf{A^{\mathit{j}}}^{H}\mathbf{A^{\mathit{j}}}\bigr\}$,
can be computed in $\Theta\left(N^{2.375}\right)$ time by resorting
to the Coppersmith-Winograd algorithm \cite{Coppersmith-Winograd}.
Therefore, the total computational complexity of the IFSLA-VBI is
\textcolor{blue}{$\Theta\left(2N^{2}T+N^{2}M+BN^{2.375}\right)$ }per
iteration. Both the Turbo-CS and Turbo-VBI algorithms contain a large-scale
matrix inverse in each iteration. Therefore, the computational complexity
of the Turbo-CS and Turbo-VBI is $\Theta\left(N^{3}\right)$ per iteration.

\section{Extension to Structured Sparse Priors\label{sec:Extension-to-Structured}}

\begin{figure}[t]
\begin{centering}
\includegraphics[width=80mm]{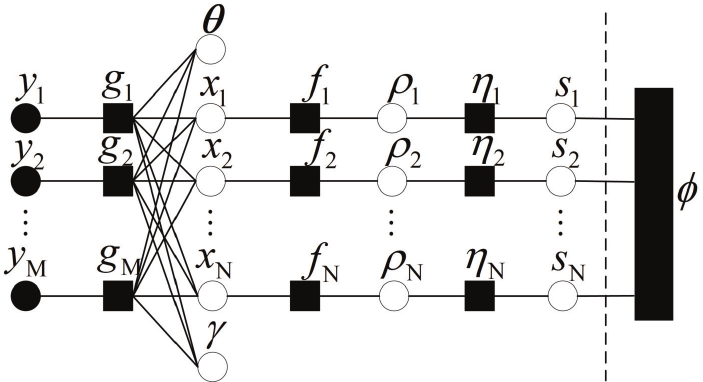}
\par\end{centering}
\caption{\label{fig:factor graph}Factor graph of the joint distribution $p\left(\boldsymbol{v},\boldsymbol{y}\right)$.}
\end{figure}
\textcolor{blue}{}
\begin{table}[t]
\caption{\label{tab:Factors_Table}Factors, Distributions and Functional forms
in Fig. \ref{fig:factor graph}. $\mathbf{F}_{m}\left(\boldsymbol{\theta}\right)$
denotes the $m\textrm{-th}$ row of $\mathbf{F}\left(\boldsymbol{\theta}\right)$\textcolor{blue}{.}}

\centering{}%
\begin{tabular}{|c|c|c|}
\hline 
Factor & Distribution & Functional form\tabularnewline
\hline 
\hline 
$g_{m}$ & $p\left(y_{m}\mid\boldsymbol{x},\gamma,\boldsymbol{\theta}\right)$ & $\mathcal{CN}\left(y_{m};\mathbf{F}_{m}\left(\boldsymbol{\theta}\right)\boldsymbol{x},\gamma^{-1}\right)$\tabularnewline
\hline 
$f_{n}$ & $p\left(x_{n}\mid\rho_{n}\right)$ & $\mathcal{CN}\left(x_{n};0,\rho_{n}^{-1}\right)$\tabularnewline
\hline 
$\eta_{n}$ & $p\left(\rho_{n}\mid s_{n}\right)$ & $\begin{cases}
\textrm{Ga}\left(\rho_{n};a_{n},b_{n}\right), & s_{n}=1\\
\textrm{Ga}\left(\rho_{n};\overline{a}_{n},\overline{b}_{n}\right), & s_{n}=0
\end{cases}$\tabularnewline
\hline 
\textcolor{blue}{$\phi$} & \textcolor{blue}{$p\left(\boldsymbol{s};\epsilon\right)$} & \textcolor{blue}{depends on application}\tabularnewline
\hline 
\end{tabular}
\end{table}
For many practical applications in wireless communications, the sparse
signal $\boldsymbol{x}$ usually has complicated sparse structures.
For example, in our considered scenario for 6G-based target localization
(Subsection \ref{subsec:5G-based-Target-Detection}), a large target
can be viewed as a cluster of target points \cite{Huangzhe_TurboSBI}.
In this case, the non-zero elements of the position-domain channel
vector are concentrated on a few bursts. And thus, the position-domain
channel exhibits a 2-D burst sparsity \textcolor{blue}{\cite{Xu_ISAC_MRF}}.
By exploiting this structured sparsity, we can further enhance the
performance of target localization. Some recent works also exploited
different structured sparsities during algorithm design. In \cite{LiuAn_TurboOAMP}\cite{LiuAn_CE_Turbo_CS},
the authors used a Markov chain model to describe the burst sparsity
of the angular-domain channel and improved the performance of channel
estimation significantly. Besides, the authors in \cite{Gao_common_sparsity,Lian_dynamic_sparsity}
exploited the temporal correlation of the support set of angular-domain
channels. Motivated by these, it is essential to extend our proposed
IFSLA-VBI algorithm from an independent sparse prior to more complicated
structured sparse priors.

\subsection{Turbo-IFSLA-VBI Algorithm}

Consider a more general distribution for the support vector, denoted
by $p\left(\boldsymbol{s};\epsilon\right)$, where $\epsilon$ denotes
the prior parameters. We can choose a proper $p\left(\boldsymbol{s};\epsilon\right)$
to model different sparse structures in practical applications. The
factor graph of the joint distribution $p\left(\boldsymbol{v},\boldsymbol{y}\right)$
is shown in Fig. \ref{fig:factor graph}, while the associated factor
nodes are listed in Table \ref{tab:Factors_Table}. Due to the complicated
internal structure of $\boldsymbol{s}$, the factor graph usually
contains loops. In this case, exact Bayesian inference is known to
be NP-hard \cite{Cooper_NP}.
\begin{figure}[t]
\begin{centering}
\includegraphics[width=80mm]{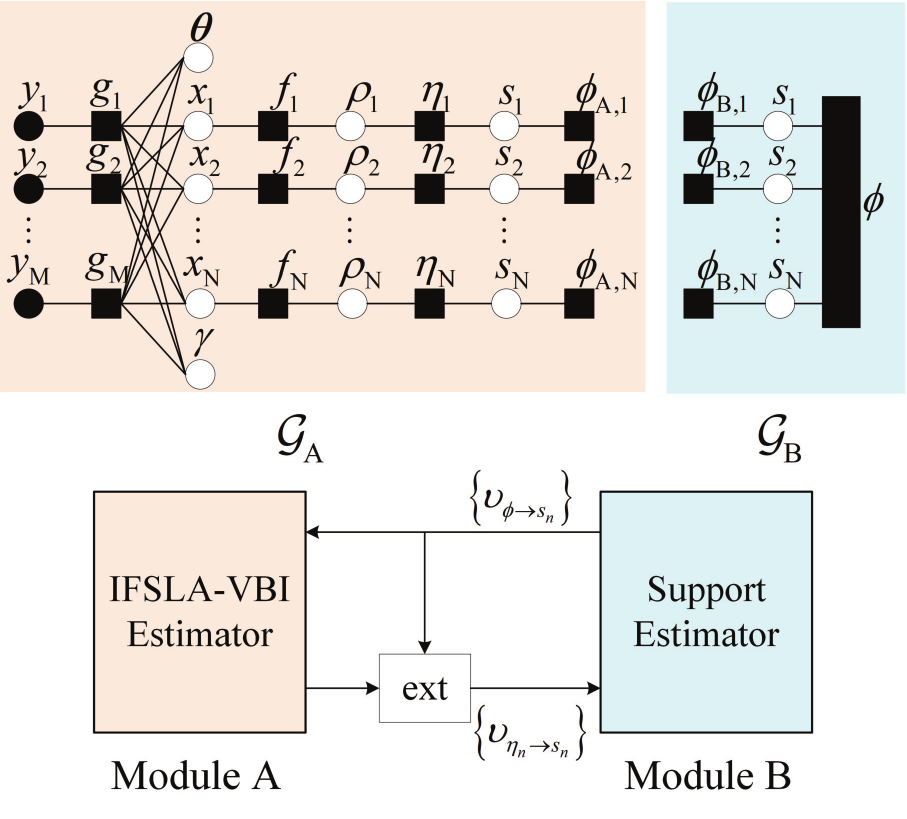}
\par\end{centering}
\caption{\label{fig:Turbo-Algorithm-decoupled-graph}The turbo approach yields
a decoupled factor graph and the framework of the Turbo-IFSLA-VBI
algorithm.}
\end{figure}

Inspired by the turbo approach \cite{Som_TurboAMP}, we propose a
Turbo-IFSLA-VBI algorithm by combining the IFSLA-VBI estimator with
message passing. We first partition the factor graph in Fig. \ref{fig:factor graph}
along the dash line into two decoupled subgraphs, denoted by $\mathcal{G}_{\mathrm{A}}$
and $\mathcal{G}_{\mathrm{B}}$, respectively, as shown in Fig. \ref{fig:Turbo-Algorithm-decoupled-graph}.
To be more specific, $\mathcal{G}_{\mathrm{A}}$ describes the internal
structure of hidden variables with an independent sparse prior, while
$\mathcal{G}_{\mathrm{B}}$ describes the more complicated internal
structure of the support vector. Then we design Module A and Module
B to perform Bayesian inference over the two subgraphs, respectively.
For $\mathcal{G}_{\mathrm{A}}$, we adopt the proposed IFSLA-VBI estimator
to compute each variational distribution approximately. For $\mathcal{G}_{\mathrm{B}}$,
we perform message passing to compute the marginal posterior of $\boldsymbol{s}$.
\textcolor{blue}{The two modules need to work alternately and exchange
extrinsic messages until converge to a stationary point. And the output
messages of one module form the priors for another module. Specifically,
the extrinsic messages from Module A to B are denoted by $\left\{ \upsilon_{\eta_{n}\rightarrow s_{n}}\right\} $,
while the extrinsic messages from Module B to A are denoted by $\left\{ \upsilon_{\phi\rightarrow s_{n}}\right\} $.
}Formally, we define two turbo-iteration factor nodes:\textcolor{blue}{
\begin{equation}
\begin{aligned}\phi_{\mathrm{A},n}\left(s_{n}\right) & \triangleq\upsilon_{\phi\rightarrow s_{n}}\left(s_{n}\right),n=1,\ldots,N,\\
\phi_{\mathrm{B},n}\left(s_{n}\right) & \triangleq\upsilon_{\eta_{n}\rightarrow s_{n}}\left(s_{n}\right),n=1,\ldots,N,
\end{aligned}
\label{eq:likelihood_message}
\end{equation}
where $\phi_{\mathrm{A},n}\left(s_{n}\right)$ can be viewed as the
prior information for Module A.} \textcolor{blue}{And for each turbo
iteration, the extrinsic message $\upsilon_{\eta_{n}\rightarrow s_{n}}\left(s_{n}\right)$
from Module A to B can be computed by subtracting the prior information
$\phi_{\mathrm{A},n}\left(s_{n}\right)$ from posterior information,}
\begin{equation}
\upsilon_{\eta_{n}\rightarrow s_{n}}\left(s_{n}\right)\varpropto q\left(s_{n}\right)/\phi_{\mathrm{A},n}\left(s_{n}\right),\label{eq:likelihood_moduleA}
\end{equation}
where $q\left(s_{n}\right)$ is the approximate posterior distribution
obtained in (\ref{eq:q(s)}).

\subsection{An Example: Markov Random Field for 2-D Burst Sparsity}

To elaborate on how Module B performs message passing more clearly,
we use the 6G-based target localization scenario as an example. As
discussed previously, since the targets are usually distributed in
clusters, the virtual position-domain channel exhibits a 2-D burst
sparsity. To exploit this, we introduce a Markov random field (MRF)
model \cite{book_MRF,Xu_ISAC_MRF}. The support vector is modeled
as
\begin{equation}
\begin{aligned}p\left(\boldsymbol{s}\right) & =\frac{1}{Z}\exp\left(\sum_{n=1}^{N}\left(\frac{1}{2}\sum_{i\in\mathcal{N}_{n}}\beta s_{i}-\alpha\right)s_{n}\right)\\
 & =\frac{1}{Z}\left(\stackrel[n=1]{N}{\prod}\underset{i\in\mathcal{N}_{n}}{\prod}\chi_{\beta}\left(s_{n},s_{i}\right)\right)^{\frac{1}{2}}\prod_{n=1}^{N}\chi_{\alpha}\left(s_{n}\right),
\end{aligned}
\end{equation}
where $\chi_{\alpha}\left(s_{n}\right)\triangleq\exp\left(-\alpha s_{n}\right)$,
$\chi_{\beta}\left(s_{n},s_{i}\right)\triangleq\exp\left(\beta s_{n}s_{i}\right)$,
$\mathcal{N}_{n}$ is the index set of the neighbor nodes of $s_{n}$,
and $Z$ is the partition function. The parameter $\alpha$ controls
the degree of sparsity and the parameter $\beta$ affects the size
of non-zero bursts.
\begin{figure}[t]
\begin{centering}
\includegraphics[width=70mm]{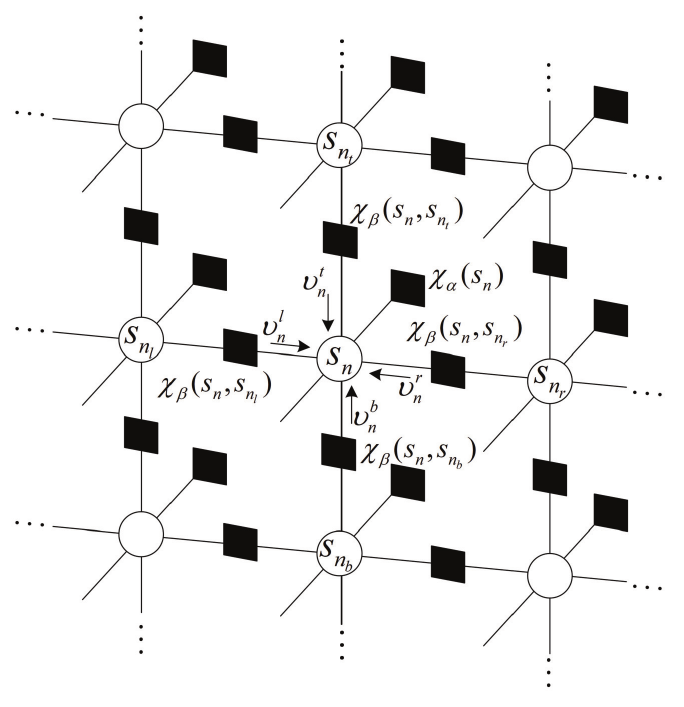}
\par\end{centering}
\caption{\label{fig:factor_MRf}The factor graph of the 4-connected MRF model.}
\end{figure}

\textcolor{blue}{We give the factor graph of the MRF model in Fig.
\ref{fig:factor_MRf}, where $\left\{ s_{n}\right\} _{n=1}^{N}$ denote
variables nodes and $\left\{ \chi_{\alpha},\chi_{\beta}\right\} $
denote factor nodes. Consider a variable node $s_{n}$, the left,
right, top, and bottom neighboring variable nodes of $s_{n}$ are
represented as $s_{n_{l}}$, $s_{n_{r}}$, $s_{n_{t}}$, and $s_{n_{b}}$,
respectively. To simplify the notation, we use $\pi_{n}^{in}$ to
abbreviate $\phi_{\mathrm{B},n}\left(s_{n}=1\right)$ for $n=1,\ldots,N$.}
In the following, we obey the sum-product rule to derive messages
over the factor graph \cite{Ksch_sum-product}.

For $s_{n}$, the input messages from the left, right, top, and bottom
neighbor nodes, denoted by $\upsilon_{n}^{l}$, $\upsilon_{n}^{r}$,
$\upsilon_{n}^{t}$, and $\upsilon_{n}^{b}$, respectively, follow
Bernoulli distributions, where $\upsilon_{n}^{l}$ is given by\textcolor{blue}{
\begin{align}
\upsilon_{n}^{l} & \propto\sum_{s_{n_{l}}}\upsilon_{\eta_{n_{l}}\rightarrow s_{n_{l}}}\prod_{k\in\left\{ l,t,b\right\} }\upsilon{}_{n_{l}}^{k}\chi_{\alpha}\left(s_{n_{l}}\right)\chi_{\beta}\left(s_{n},s_{n_{l}}\right)\nonumber \\
 & \propto\kappa_{n}^{l}\delta\left(s_{n}-1\right)+(1-\kappa_{n}^{l})\delta\left(s_{n}\right),
\end{align}
}where
\[
\kappa_{n}^{l}=\tfrac{\pi_{n_{l}}^{in}\prod_{k\in\left\{ l,t,b\right\} }\kappa_{n_{l}}^{k}e^{-\alpha+\beta}+\left(1-\pi_{n_{l}}^{in}\right)\prod_{k\in\left\{ l,t,b\right\} }\left(1-\kappa_{n_{l}}^{k}\right)e^{\alpha-\beta}}{\left(e^{\beta}+e^{-\beta}\right)\left(\pi_{n_{l}}^{in}e^{-\alpha}\prod_{k\in\left\{ l,t,b\right\} }\kappa_{q_{l}}^{k}+\left(1-\pi_{n_{l}}^{in}\right)e^{\alpha}\prod_{k\in\left\{ l,t,b\right\} }\left(1-\kappa_{n_{l}}^{k}\right)\right)}.
\]
The messages $\upsilon_{n}^{r}$, $\upsilon_{n}^{t}$, and $\upsilon_{n}^{b}$
can be calculated in a similar way.

Then the output message for $s_{n}$ (i.e., the extrinsic message
from Module B to A) can be calculated as
\begin{align}
\upsilon_{\phi\rightarrow s_{n}} & \propto\Pi_{k\in\left\{ l,r,t,b\right\} }\upsilon_{n}^{k}\chi_{\alpha}\left(s_{n_{l}}\right)\nonumber \\
 & \propto\pi_{n}^{out}\delta\left(s_{n}-1\right)+\left(1-\pi_{n}^{out}\right)\delta\left(s_{n}\right)
\end{align}
where
\[
\pi_{n}^{out}=\frac{e^{-\alpha}\Pi_{k\in\left\{ l,r,t,b\right\} }\kappa_{n}^{k}}{e^{-\alpha}\Pi_{k\in\left\{ l,r,t,b\right\} }\kappa_{n}^{k}+e^{\alpha}\Pi_{k\in\left\{ l,r,t,b\right\} }\left(1-\kappa_{n}^{k}\right)}.
\]

\subsection{Summary of the Turbo-IFSLA-VBI}

We summarize the Turbo-IFSLA-VBI algorithm in Algorithm \ref{Turbo-IF-LAVBI}.
Since the message passing is usually linear complexity, the additional
computational overhead caused by Module B is almost negligible.
\begin{algorithm}[t]
\begin{singlespace}
{\small{}\caption{\label{Turbo-IF-LAVBI}Turbo-IFSLA-VBI algorithm}
}{\small\par}

\textbf{Input:} $\boldsymbol{y}$, initial grid $\overline{\boldsymbol{\theta}}$,
iteration number $I=I_{1}+I_{2}$ ($I_{2}=I_{\textrm{A}}I_{\textrm{B}}$),
local iteration number $T$.

\textbf{Output:} $\boldsymbol{x}^{*}$, $\boldsymbol{s}^{*}$, and
$\boldsymbol{\theta}^{*}$.

\begin{algorithmic}[1]

\STATE \textbf{\% Stage 1: Initialization}

\STATE Same as the initialization stage of the IFSLA-VBI.

\STATE \textbf{\% Stage 2: Turbo-IFSLA-VBI Algorithm}

\FOR{${\color{blue}{\color{black}i_{b}=1,\cdots,I_{\textrm{B}}}}$}

\STATE \textbf{\% Module A: IFSLA-VBI Estimator}

\STATE Initialize $i_{a}=1$.

\WHILE{not converge and $i_{a}\leq I_{\textrm{A}}$}

\STATE Optimize $q\left(\boldsymbol{x}\right)$, $q\left(\boldsymbol{\rho}\right)$,
$q\left(\boldsymbol{s}\right)$, $q\left(\gamma\right)$, and \textbf{$q\left(\boldsymbol{\theta}\right)$}
alternatively according to the IFSLA-VBI algorithm.

\STATE $i_{a}=i_{a}+1$.

\ENDWHILE

\STATE Compute $\left\{ \upsilon_{\eta_{n}\rightarrow s_{n}}\right\} $
based on (\ref{eq:likelihood_moduleA}) and send it to Module B.

\STATE \textbf{\% Module B: Support Estimator}

\STATE Perform message passing over $\mathcal{G}_{\mathrm{B}}$,
send $\left\{ \upsilon_{\phi\rightarrow s_{n}}\right\} $ to Module
A.

\ENDFOR

\STATE Output $\boldsymbol{x}^{*}=\boldsymbol{\mu}_{x}$, $s_{n}^{*}=\widetilde{\lambda}_{n},\forall n$,
and $\mathbf{\boldsymbol{\theta}^{\mathit{j}}}^{*}=\boldsymbol{\mu}_{\theta^{j}},\forall j$.

\end{algorithmic}
\end{singlespace}
\end{algorithm}

\section{Simulation Results}

In this section, we apply the proposed algorithm to solve two practical
application problems and verify its advantages compared to other baselines.
Different algorithms are elaborated below.
\begin{table*}[t]
\centering{}\caption{\label{tab:Complexity}Complexity order, numerical value of order,
and CPU time for different algorithms}
\begin{tabular}{|c|c|c|c|c|c|}
\hline 
\multirow{2}{*}{Algorithms} & \multirow{2}{*}{Complexity order} & \multicolumn{2}{c|}{Numerical value of order} & \multicolumn{2}{c|}{CPU time}\tabularnewline
\cline{3-6} \cline{4-6} \cline{5-6} \cline{6-6} 
 &  & Application 1 & Application 2 & Application 1 & Application 2\tabularnewline
\hline 
\hline 
OGSBI & $\Theta\left(IN^{3}\right)$ & $\Theta\left(1.0\times10^{8}\right)$ & $\Theta\left(6.7\times10^{9}\right)$ & 0.0561s & \tabularnewline
\hline 
SBL/EM & $\Theta\left(I_{out}I_{in}N^{3}+I_{out}N^{2}M\right)$ & $\Theta\left(1.1\times10^{9}\right)$ & $\Theta\left(7.4\times10^{10}\right)$ & 0.433s & \textcolor{blue}{25.3s}\tabularnewline
\hline 
Turbo-CS/EM & $\Theta\left(I_{out}I_{in}N^{3}+I_{out}N^{2}M\right)$ & $\Theta\left(1.1\times10^{9}\right)$ & $\Theta\left(7.4\times10^{10}\right)$ & 0.449s & 18.4s\tabularnewline
\hline 
Turbo-VBI/EM & $\Theta\left(I_{out}I_{in}N^{3}+I_{out}N^{2}M\right)$ & $\Theta\left(1.1\times10^{9}\right)$ & $\Theta\left(7.4\times10^{10}\right)$ & 0.453s & 25.4s\tabularnewline
\hline 
SLA-VBI & \textcolor{blue}{$\Theta\left(I\left(2N^{3}+N^{2}M+BN^{2.375}\right)\right)$} & $\Theta\left(2.6\times10^{8}\right)$ & $\Theta\left(2.0\times10^{10}\right)$ & 0.185s & \tabularnewline
\hline 
IFSLA-VBI & \textcolor{blue}{$\Theta\left(I\left(2N^{2}T+N^{2}M+BN^{2.375}\right)\right)$} & $\Theta\left(7.4\times10^{7}\right)$ & $\Theta\left(7.2\times10^{9}\right)$ & 0.0550s & 2.00s\tabularnewline
\hline 
Turbo-IFSLA-VBI & \textcolor{blue}{$\Theta\left(I\left(2N^{2}T+N^{2}M+BN^{2.375}\right)\right)$} & $\Theta\left(7.4\times10^{7}\right)$ & $\Theta\left(7.2\times10^{9}\right)$ &  & 2.02s\tabularnewline
\hline 
\end{tabular}
\end{table*}
\begin{figure*}[t]
\centering{}%
\begin{minipage}[t]{0.45\textwidth}%
\begin{center}
\includegraphics[clip,width=70mm]{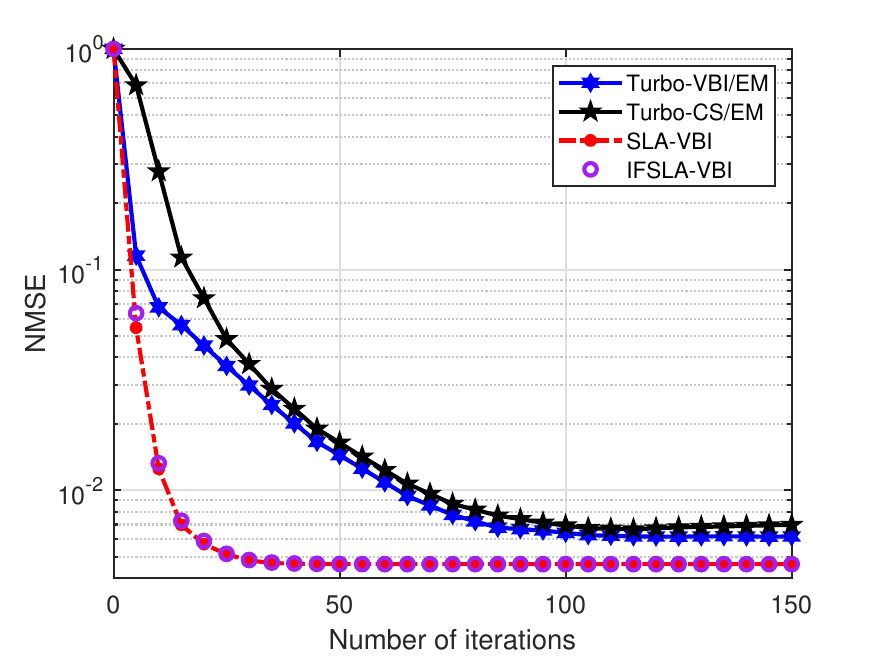}
\par\end{center}
\caption{\textcolor{blue}{\label{fig:Ap1-convergence-behavior}Convergence
behavior of the Turbo-VBI, Turbo-CS, SLA-VBI, and IFSLA-VBI. The number
of pilot sequences is set to $64$ and $\textrm{SNR}$ is 10~dB.}}
\end{minipage}\hfill{}%
\begin{minipage}[t]{0.45\textwidth}%
\begin{center}
\includegraphics[clip,width=70mm]{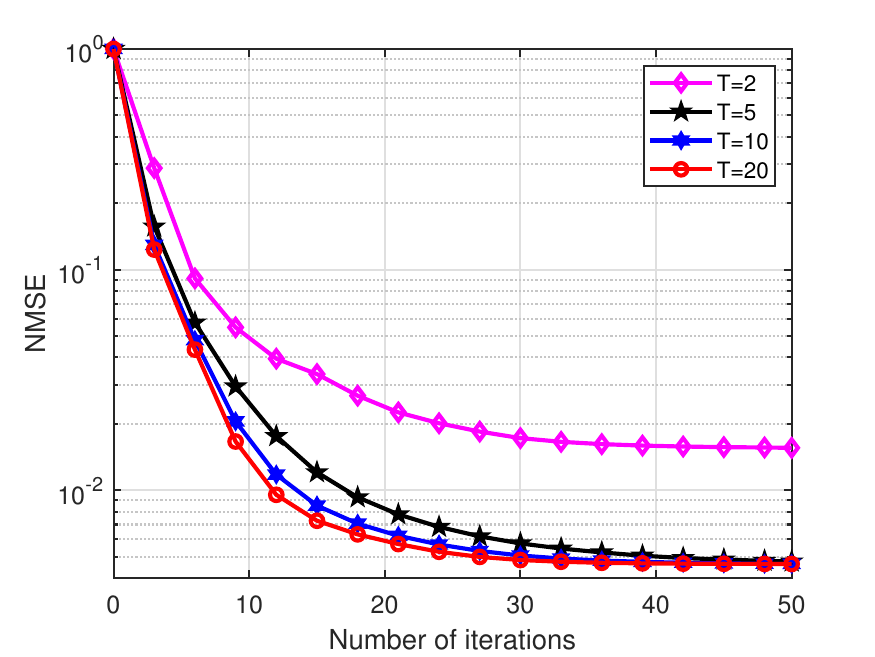}
\par\end{center}
\caption{\textcolor{blue}{\label{fig:Ap1-convergence_versus_T}}Convergence
behavior of the IFSLA-VBI algorithm with different local iteration
numbers. We set $T=2,5,10,20$.}
\end{minipage}
\end{figure*}

\begin{itemize}
\item \textbf{OGSBI \cite{OGSBI}: }It is a single-loop algorithm based
on linear approximation.
\item \textbf{EM-based sparse Bayesian learning (SBL) \cite{Tipping_SBL}:}
It is a double-loop EM framework, where the E-step applies a SBL estimator
to recover sparse signals and the M-step uses a gradient ascent method
to update dynamic grid parameters.
\item \textbf{EM-based Turbo-CS }\cite{Yuan_TurboCS,LiuAn_TurboOAMP,Huangzhe_TurboSBI}\textbf{:}
It is a double-loop EM framework, where the E-step is the Turbo-CS
algorithm and the M-step performs gradient ascent update.
\item \textbf{EM-based Turbo-VBI }\cite{LiuAn_CE_Turbo_VBI,LiuAn_directloc_vehicles}\textbf{:}
It is a double-loop EM framework, where the E-step is the Turbo-VBI
algorithm and the M-step performs gradient ascent update.
\item \textbf{SLA-VBI: }The proposed SLA-VBI algorithm is single-loop but
involves two complicated matrix inverse operations in each iteration.
\item \textbf{IFSLA-VBI: }It is the simplified version of the SLA-VBI, where
the matrix inverse is approximated by the MM framework.
\item \textbf{Turbo-IFSLA-VBI: }It is the extension of the IFSLA-VBI, which
can exploit different sparse structures.
\end{itemize}
For the OGSBI and our proposed algorithms, the maximum number of iterations
is set to $I=50$. For the double-loop EM-based algorithms, the inner
iteration number of the Bayesian estimator is set to $I_{in}=10$
and the outer iteration number of the EM is set to $I_{out}=50$.
For the IFSLA-VBI and Turbo-IFSLA-VBI, the number of iterations used
to approximate the matrix inverse is set to $T=10$. As seen from
Table \ref{tab:Complexity}, both the complexity order and the average
run time of the proposed IFSLA-VBI are significantly lower than the
double-loop EM-based methods \footnote{Note that we measure the average run time via MATLAB on a laptop computer
with a $2.5\ \textrm{GHz}$ CPU.}. \textcolor{blue}{The source code is available at https://github.com/ZJU-XWK/SLA-VBI.}

\subsection{Massive MIMO Channel Estimation}

\subsubsection{Implementation Details}

In the simulations, the BS is equipped with a ULA of $128$ antennas.
The pilot symbols are generated with random phase under unit power
constrains. The number of AoD grid points is set to $128$. For the
three-layer sparse prior model, we set $a_{n}=1$, $b_{n}=1$, $\overline{a}_{n}=1$,
$\overline{b}_{n}=10^{-5}$, $c=10^{-6}$, and $d=10^{-6}$. We choose
the normalized mean square error (NMSE) as the performance metric
for channel estimation. \textcolor{blue}{The related parameters for
simulations are listed in Table \ref{tab:Related-parameter}. }
\begin{figure}[t]
\begin{centering}
\includegraphics[width=70mm]{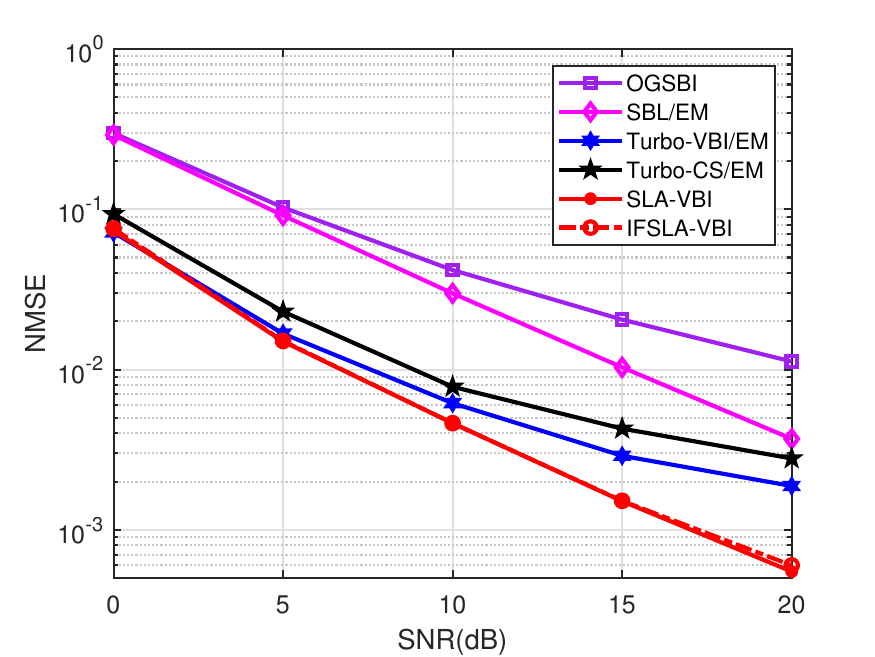}
\par\end{centering}
\caption{\label{fig:Ap1-NMSE_versus_SNR}The NMSE performance of the channel
estimation versus SNR. The number of pilot sequences is $64$.}
\end{figure}
\begin{figure}[t]
\begin{centering}
\includegraphics[width=70mm]{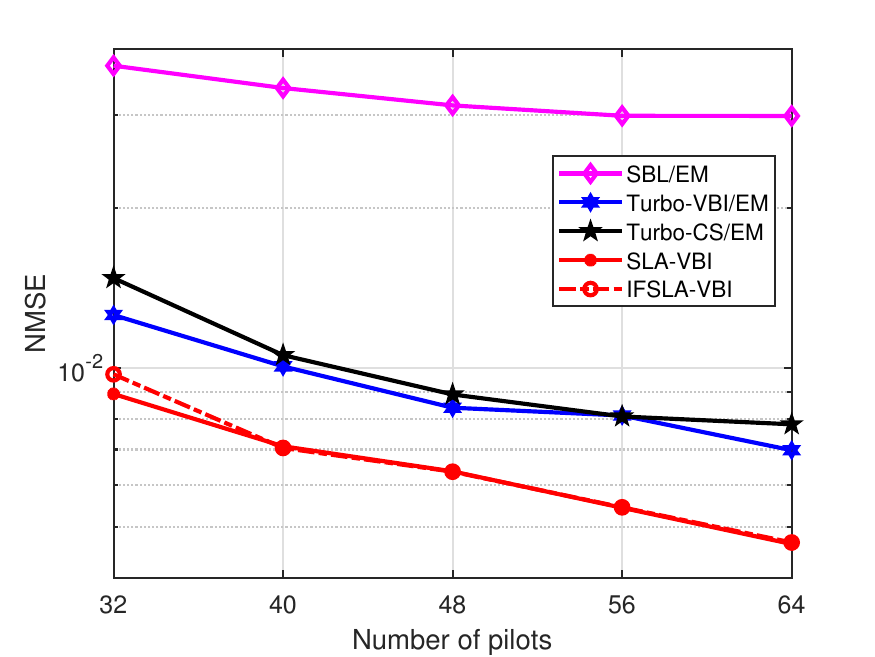}
\par\end{centering}
\caption{\label{fig:Ap1-NMSE_versus_CSrate}The NMSE performance of the channel
estimation versus number of pilot sequences. We set $\textrm{SNR}=10~\textrm{dB}$.}
\end{figure}

\subsubsection{Convergence Behavior}

In Fig. \ref{fig:Ap1-convergence-behavior}, we compare the convergence
behavior of different algorithms. The proposed IFSLA-VBI has a much
faster convergence speed than the double-loop EM-based Turbo-CS and
Turbo-VBI. Besides, the IFSLA-VBI has similar convergence behavior
to the more complicated SLA-VBI, which reflects that the matrix inverse
approximated by the local iterations is accurate enough. In Fig. \ref{fig:Ap1-convergence_versus_T},
we change the number of local iterations used to approximate the matrix
inverse and evaluate the convergence behavior of the IFSLA-VBI algorithm.
As can be seen, the IFSLA-VBI still works well for $T$ as small as
$5$. That is to say, the IFSLA-VBI can achieve comparable performance
to the SLA-VBI while greatly reducing the computational overhead.

\subsubsection{Influence of SNR}

\textcolor{blue}{In Fig. \ref{fig:Ap1-NMSE_versus_SNR}, we show the
performance of channel estimation versus SNR. }It can be seen that
the performance of all the algorithms improves as the SNR increases.
With limited training sequences, the OGSBI works poorly. Besides,
the EM-based Turbo-VBI works better than the EM-based SBL, which reflects
the advantage of the three-layer sparse prior model used in Turbo-VBI.
Furthermore, the proposed IFSLA-VBI can achieve a significant performance
gain over the state-of-the-art Turbo-CS and Turbo-VBI, especially
in the high SNR regions. This is because the proposed IFSLA-VBI can
output Bayesian estimation of both sparse signals and grid parameters
but the EM-based methods can only provide a point estimation of grid
parameters. Finally, the curves of the IFSLA-VBI and SLA-VBI almost
overlap, i.e., they have almost the same performance.

\subsubsection{Influence of Number of Pilots}

\textcolor{blue}{In Fig. \ref{fig:Ap1-NMSE_versus_CSrate}, we focus
on how the number of pilot sequences affects the performance of channel
estimation.} As the number of pilot sequences increases, the performance
of all the algorithms improves. And it is obvious that the proposed
IFSLA-VBI still performs better than the EM-based methods.
\begin{table}[t]
\textcolor{blue}{\caption{\label{tab:Related-parameter}Related parameters for simulations}
}
\centering{}\textcolor{blue}{}%
\begin{tabular}{|c|c|c|c|}
\hline 
\multicolumn{2}{|c|}{\textcolor{blue}{Application 1}} & \multicolumn{2}{c|}{\textcolor{blue}{Application 2}}\tabularnewline
\hline 
\textcolor{blue}{Parameter} & \textcolor{blue}{Value} & \textcolor{blue}{Parameter} & \textcolor{blue}{Value}\tabularnewline
\hline 
\hline 
\textcolor{blue}{Antenna number} & \textcolor{blue}{128} & \textcolor{blue}{Antenna number} & \textcolor{blue}{64}\tabularnewline
\hline 
\textcolor{blue}{Pilot number} & \textcolor{blue}{64} & \textcolor{blue}{RF chain number} & \textcolor{blue}{16}\tabularnewline
\hline 
\textcolor{blue}{AoD grid points} & \textcolor{blue}{128} & \textcolor{blue}{Subcarrier number} & \textcolor{blue}{1024}\tabularnewline
\hline 
 &  & \textcolor{blue}{Subcarrier interval} & \textcolor{blue}{$30~\textrm{kHz}$}\tabularnewline
\hline 
 &  & \textcolor{blue}{Position grid points} & \textcolor{blue}{512}\tabularnewline
\hline 
\end{tabular}
\end{table}
\begin{figure}[t]
\subfloat[\label{fig:The-2-D-platform.}The 2-D platform.]{\centering{}\includegraphics[width=45mm]{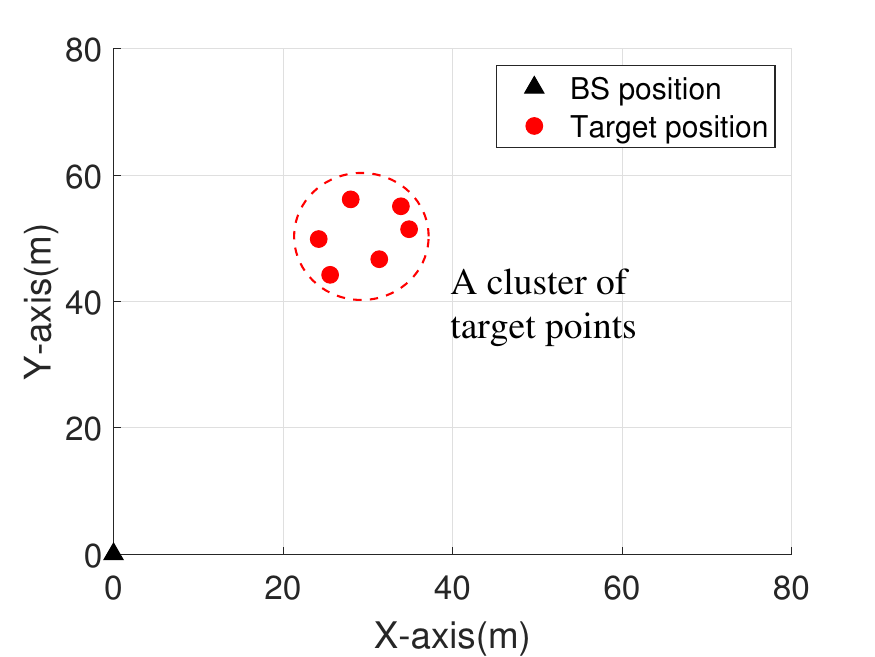}}\subfloat[\label{fig:Target-localization-via}Target localization result.]{\centering{}\includegraphics[width=45mm]{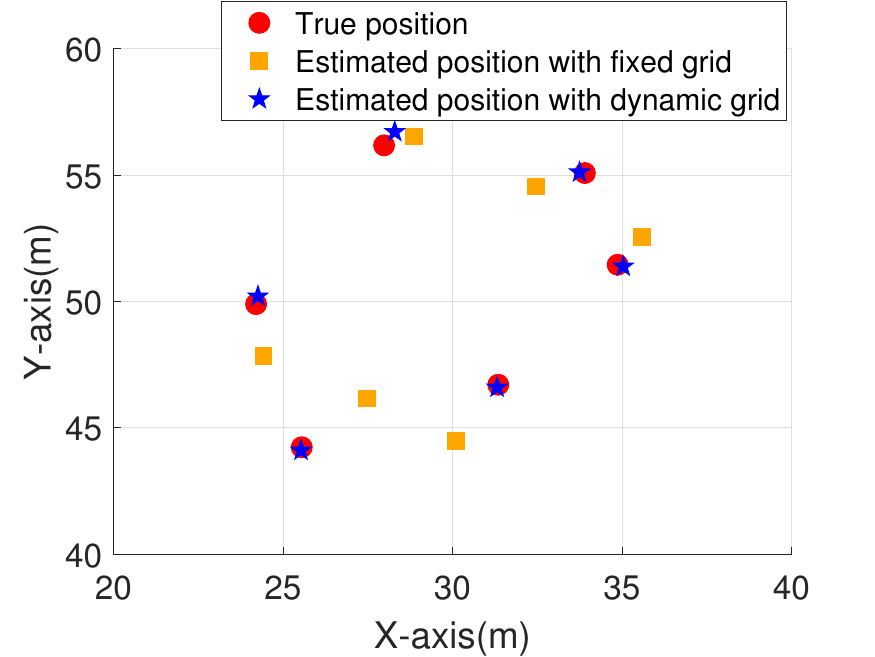}}\caption{Illustration of the simulation setup for 6G-based target localization.}
\end{figure}
\begin{figure}[t]
\begin{centering}
\includegraphics[width=70mm]{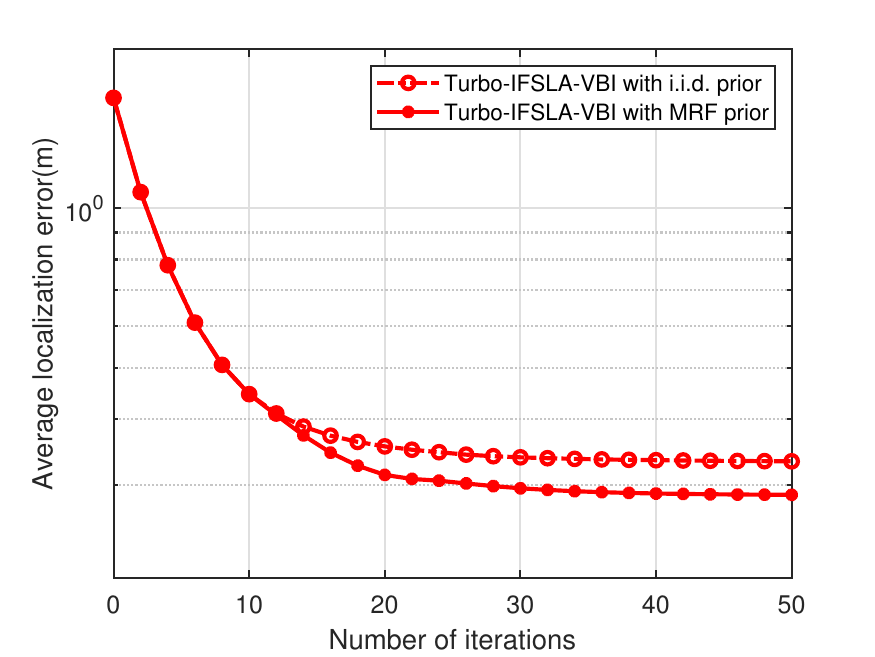}
\par\end{centering}
\caption{\textcolor{blue}{\label{fig:HBG-Convergence-behavior}Convergence
behavior of the Turbo-IFSLA-VBI with an i.i.d. prior and the MRF prior.
We set $\textrm{SNR}=0~\textrm{dB}$.}}
\end{figure}
\begin{figure}[t]
\begin{centering}
\includegraphics[width=70mm]{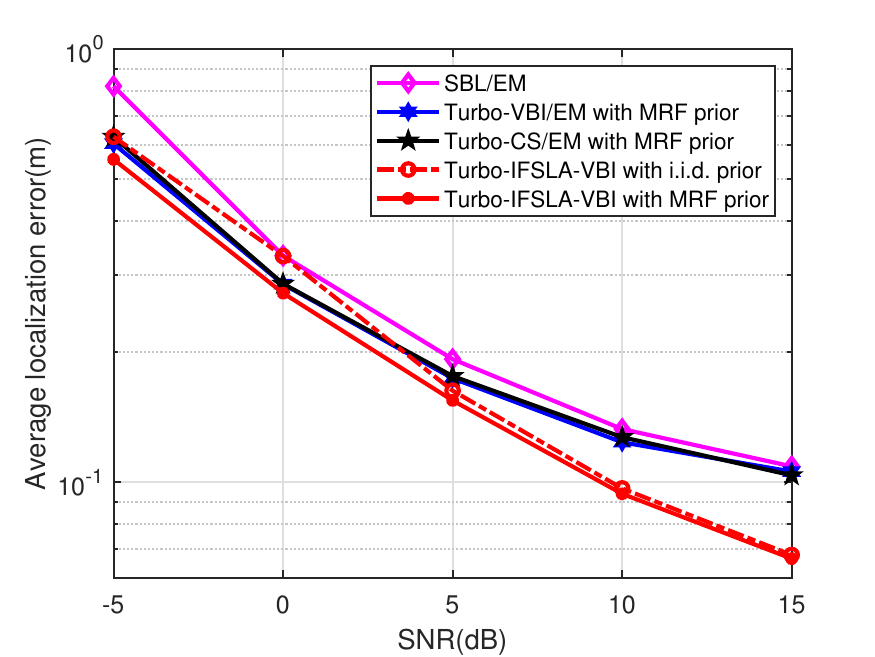}
\par\end{centering}
\caption{\label{fig:HBG-RMSE-SNR}\textcolor{blue}{Average localization error
versus SNR. The number of RF chains is $16$.}}

\begin{centering}
\includegraphics[width=70mm]{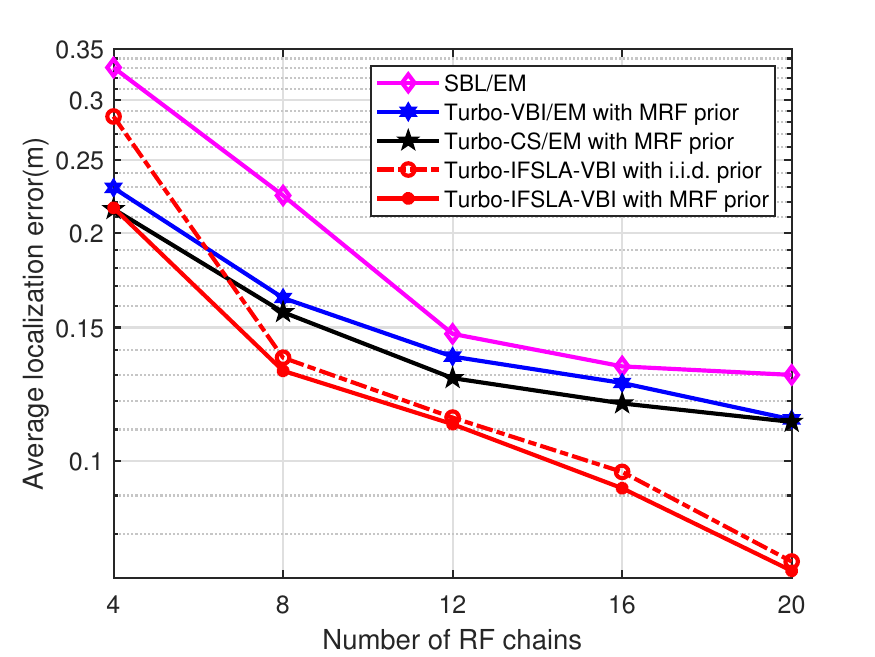}
\par\end{centering}
\caption{\label{fig:HBG-RMSE-RFchains}\textcolor{blue}{Average localization
error versus number of RF chains. We set $\textrm{SNR}=10~\textrm{dB}$}.}
\end{figure}

\subsection{6G-based Target Localization}

\subsubsection{Implementation Details}

We consider a 2-D platform, where the BS is deployed at the corner
with the coordinates $\left(0~\textrm{m},0~\textrm{m}\right)$ and
the targets are concentrated in a cluster, as shown in Fig. \ref{fig:The-2-D-platform.}.
The BS has $64$ antennas and $16$ RF chains. The number of OFDM
subcarriers is set to $1024$ and the subcarrier interval is $30~\textrm{kHz}$.
Probing signals are generated with random phase under unit power constrains,
\textcolor{black}{and they are inserted at intervals of $32$ OFDM
subcarriers, i.e., $\left|\mathcal{N}_{b}\right|=\tfrac{1024}{32}=32$}.
The RF combining matrix $\mathbf{W}_{\textrm{RF}}$ is partially orthogonal.
The model parameters of the MRF is set to $\alpha=0.3$ and $\beta=0.5$.
We introduce a position-domain dynamic grid with $512$ grid points
for target localization. \textcolor{blue}{The simulation parameters
are listed in Table \ref{tab:Related-parameter}.} The dynamic grid
will greatly improve localization accuracy, as shown in Fig. \ref{fig:Target-localization-via}.
In contrast, there is a glaring mismatch between the true positions
and the estimated positions when using a fixed sampling grid. 

\subsubsection{Convergence Behavior}

In Fig. \ref{fig:HBG-Convergence-behavior}, we compare the convergence
behavior of the proposed Turbo-IFSLA-VBI algorithm with different
sparse priors. It can be seen that the algorithm with the MRF prior
has a much smaller localization error after convergence. This indicates
that the proposed Turbo-IFSLA-VBI has the ability to utilize the structured
sparse prior information to improve the performance. 

\subsubsection{Influence of SNR}

In Fig. \ref{fig:HBG-RMSE-SNR}, we evaluate the performance of target
localization versus SNR. The proposed Turbo-IFSLA-VBI with the MRF
prior has the smallest localization error among all the algorithms.
There is a significant performance gap between the Turbo-IFSLA-VBI
with the MRF prior and the same algorithm with an i.i.d. prior, which
reflects that the MRF prior can fully exploit the 2-D burst sparsity
of the position-domain channel. In the high SNR regions, the localization
error of the EM-based methods is much larger than our proposed algorithm,
which reflects the advantage of Bayesian estimation of grid parameters.

\subsubsection{Influence of Number of RF Chains}

In Fig. \ref{fig:HBG-RMSE-RFchains}, we evaluate the performance
of target localization versus number of RF chains. Again, the proposed
Turbo-IFSLA-VBI with the MRF prior achieves the best performance.

\section{Conclusion}

We propose a novel SLA-VBI algorithm to recover a structured sparse
signal from a linear model with uncertain grid parameters in the sensing
matrix. In contrast to conventional EM-based methods, our proposed
algorithm can provide approximate posterior distribution of both sparse
signals and dynamic grid parameters. To reduce the computational overhead
caused by the matrix inverse, we design an inverse-free algorithm
(i.e., IFSLA-VBI) based on the MM framework. And then we extend the
proposed algorithm from an independent sparse prior to more complicated
structured sparse priors by using the turbo approach. Finally, we
apply our proposed algorithm to solve two practical applications,
i.e., massive MIMO channel estimation and 6G-based target localization.
The simulations verify that our proposed algorithm can achieve faster
convergence, lower complexity, and better performance compared to
the state-of-the-art EM-based methods.

\appendix

\subsection{\textcolor{blue}{Derivation of (\ref{eq:lnq(theta_j_final)}) - (\ref{eq:theta_post})\label{subsec:Derivation-of-equations}}}

\textcolor{blue}{Substituting (\ref{eq:linearization_matrix}) into
(\ref{eq:q(theta)}) and ignoring the terms that are not related to
$\boldsymbol{\theta}^{j}$, $q\left(\boldsymbol{\theta}^{j}\right)$
can be derived as
\begin{align}
\ln q\left(\boldsymbol{\theta}^{j}\right)\propto & \left\langle \ln p\left(\boldsymbol{y}\mid\boldsymbol{x},\boldsymbol{\theta},\gamma\right)\right\rangle _{q\left(\boldsymbol{x}\right)q\left(\gamma\right)\Pi_{i\neq j}q\left(\boldsymbol{\theta}^{i}\right)}+\ln p\left(\boldsymbol{\theta}^{j}\right)\nonumber \\
\propto & -\left\langle \gamma\right\rangle \left\langle \left\Vert \boldsymbol{y}-\overline{\mathbf{F}}\left(\boldsymbol{\theta}\right)\boldsymbol{x}\right\Vert ^{2}\right\rangle _{q\left(\boldsymbol{x}\right)\Pi_{i\neq j}q\left(\boldsymbol{\theta}^{i}\right)}\nonumber \\
 & +\ln\mathcal{N}\left(\boldsymbol{\theta}^{j};\overline{\boldsymbol{\theta}}^{j},1/\kappa^{j}\mathbf{I}_{N}\right)\nonumber \\
\propto & -\left\langle \gamma\right\rangle \left\langle \left\Vert \boldsymbol{y}-\overline{\mathbf{F}}\left(\boldsymbol{\theta}\right)\boldsymbol{\mu}_{x}\right\Vert ^{2}\right\rangle _{\Pi_{i\neq j}q\left(\boldsymbol{\theta}^{i}\right)}\nonumber \\
 & -\left\langle \gamma\right\rangle \left\langle \textrm{Tr}\left(\overline{\mathbf{F}}\left(\boldsymbol{\theta}\right)\boldsymbol{\Sigma}_{x}\overline{\mathbf{F}}\left(\boldsymbol{\theta}\right)^{H}\right)\right\rangle _{\Pi_{i\neq j}q\left(\boldsymbol{\theta}^{i}\right)}\nonumber \\
 & +\ln\mathcal{N}\left(\boldsymbol{\theta}^{j};\overline{\boldsymbol{\theta}}^{j},1/\kappa^{j}\mathbf{I}_{N}\right)\nonumber \\
\propto & -\left\langle \gamma\right\rangle \boldsymbol{\mu}_{x}^{H}\left\langle \overline{\mathbf{F}}\left(\boldsymbol{\theta}\right)^{H}\overline{\mathbf{F}}\left(\boldsymbol{\theta}\right)\right\rangle _{\Pi_{i\neq j}q\left(\boldsymbol{\theta}^{i}\right)}\boldsymbol{\mu}_{x}\nonumber \\
 & +2\left\langle \gamma\right\rangle \mathfrak{Re}\left\{ \boldsymbol{\mu}_{x}^{H}\left\langle \overline{\mathbf{F}}\left(\boldsymbol{\theta}\right)^{H}\right\rangle _{\Pi_{i\neq j}q\left(\boldsymbol{\theta}^{i}\right)}\boldsymbol{y}\right\} \nonumber \\
 & -\left\langle \gamma\right\rangle \textrm{Tr}\left(\left\langle \overline{\mathbf{F}}\left(\boldsymbol{\theta}\right)^{H}\overline{\mathbf{F}}\left(\boldsymbol{\theta}\right)\right\rangle _{\Pi_{i\neq j}q\left(\boldsymbol{\theta}^{i}\right)}\boldsymbol{\Sigma}_{x}\right)\nonumber \\
 & +\ln\mathcal{N}\left(\boldsymbol{\theta}^{j};\overline{\boldsymbol{\theta}}^{j},1/\kappa^{j}\mathbf{I}_{N}\right),\label{eq:ln(qthetaj_middle)}
\end{align}
The first term in (\ref{eq:ln(qthetaj_middle)}) can be computed as
\begin{align}
\textrm{Term1}= & -\left\langle \gamma\right\rangle \boldsymbol{\mu}_{x}^{H}\left\langle \overline{\mathbf{F}}\left(\boldsymbol{\theta}\right)^{H}\overline{\mathbf{F}}\left(\boldsymbol{\theta}\right)\right\rangle _{\Pi_{i\neq j}q\left(\boldsymbol{\theta}^{i}\right)}\boldsymbol{\mu}_{x}\nonumber \\
= & -2\left\langle \gamma\right\rangle \boldsymbol{\mu}_{x}^{H}\mathfrak{Re}\left\{ \textrm{diag}\left(\boldsymbol{\theta}^{j}-\hat{\boldsymbol{\mu}}_{\theta^{j}}\right)\mathbf{A^{\mathit{j}}}^{H}\mathbf{F}\left(\hat{\boldsymbol{\mu}}_{\theta}\right)\right\} \boldsymbol{\mu}_{x}\nonumber \\
 & -\left\langle \gamma\right\rangle \boldsymbol{\mu}_{x}^{H}\textrm{diag}\left(\boldsymbol{\theta}^{j}-\hat{\boldsymbol{\mu}}_{\theta^{j}}\right)\mathbf{A^{\mathit{j}}}^{H}\mathbf{A^{\mathit{j}}}\textrm{diag}\left(\boldsymbol{\theta}^{j}-\hat{\boldsymbol{\mu}}_{\theta^{j}}\right)\boldsymbol{\mu}_{x}\nonumber \\
= & -2\left\langle \gamma\right\rangle \mathfrak{Re}\left\{ \boldsymbol{\mu}_{x}^{H}\textrm{diag}\left(\boldsymbol{\theta}^{j}-\hat{\boldsymbol{\mu}}_{\theta^{j}}\right)\mathbf{A^{\mathit{j}}}^{H}\mathbf{F}\left(\hat{\boldsymbol{\mu}}_{\theta}\right)\boldsymbol{\mu}_{x}\right\} \nonumber \\
 & -\left\langle \gamma\right\rangle \mathfrak{Re}\Bigl\{\left(\boldsymbol{\theta}^{j}-\hat{\boldsymbol{\mu}}_{\theta^{j}}\right)^{T}\nonumber \\
 & \hspace{0.6cm}\left[\textrm{diag}\left(\boldsymbol{\mu}_{x}\right)^{H}\mathbf{A^{\mathit{j}}}^{H}\mathbf{A^{\mathit{j}}}\textrm{diag}\left(\boldsymbol{\mu}_{x}\right)\right]\left(\boldsymbol{\theta}^{j}-\hat{\boldsymbol{\mu}}_{\theta^{j}}\right)\Bigr\}\nonumber \\
= & -2\left\langle \gamma\right\rangle \mathfrak{Re}\left\{ \left(\boldsymbol{\theta}^{j}-\hat{\boldsymbol{\mu}}_{\theta^{j}}\right)^{T}\textrm{diag}\left(\boldsymbol{\mu}_{x}\right)^{H}\mathbf{A^{\mathit{j}}}^{H}\mathbf{F}\left(\hat{\boldsymbol{\mu}}_{\theta}\right)\boldsymbol{\mu}_{x}\right\} \nonumber \\
 & -\left\langle \gamma\right\rangle \left(\boldsymbol{\theta}^{j}-\hat{\boldsymbol{\mu}}_{\theta^{j}}\right)^{T}\nonumber \\
 & \hspace{0.6cm}\mathfrak{Re}\left\{ \left(\boldsymbol{\mu}_{x}\boldsymbol{\mu}_{x}^{H}\right)^{T}\odot\left(\mathbf{A^{\mathit{j}}}^{H}\mathbf{A^{\mathit{j}}}\right)\right\} \left(\boldsymbol{\theta}^{j}-\hat{\boldsymbol{\mu}}_{\theta^{j}}\right).\label{eq:first_term}
\end{align}
The second term in (\ref{eq:ln(qthetaj_middle)}) can be computed
as
\begin{align}
\textrm{Term2} & =2\left\langle \gamma\right\rangle \mathfrak{Re}\left\{ \boldsymbol{\mu}_{x}^{H}\left\langle \overline{\mathbf{F}}\left(\boldsymbol{\theta}\right)^{H}\right\rangle _{\Pi_{i\neq j}q\left(\boldsymbol{\theta}^{i}\right)}\boldsymbol{y}\right\} \nonumber \\
 & =2\left\langle \gamma\right\rangle \mathfrak{Re}\left\{ \boldsymbol{\mu}_{x}^{H}\textrm{diag}\left(\boldsymbol{\theta}^{j}-\hat{\boldsymbol{\mu}}_{\theta^{j}}\right)\mathbf{A^{\mathit{j}}}^{H}\boldsymbol{y}\right\} \nonumber \\
 & =2\left\langle \gamma\right\rangle \mathfrak{Re}\left\{ \left(\boldsymbol{\theta}^{j}-\hat{\boldsymbol{\mu}}_{\theta^{j}}\right)^{T}\textrm{diag}\left(\boldsymbol{\mu}_{x}\right)^{H}\mathbf{A^{\mathit{j}}}^{H}\boldsymbol{y}\right\} .\label{eq:second_term}
\end{align}
The third term in (\ref{eq:ln(qthetaj_middle)}) can be computed as
\begin{align}
\textrm{Term3=} & -\left\langle \gamma\right\rangle \textrm{Tr}\left(\left\langle \overline{\mathbf{F}}\left(\boldsymbol{\theta}\right)^{H}\overline{\mathbf{F}}\left(\boldsymbol{\theta}\right)\right\rangle _{\Pi_{i\neq j}q\left(\boldsymbol{\theta}^{i}\right)}\boldsymbol{\Sigma}_{x}\right)\nonumber \\
= & -\left\langle \gamma\right\rangle \textrm{Tr}\left(2\mathfrak{Re}\left\{ \textrm{diag}\left(\boldsymbol{\theta}^{j}-\hat{\boldsymbol{\mu}}_{\theta^{j}}\right)\mathbf{A^{\mathit{j}}}^{H}\mathbf{F}\left(\hat{\boldsymbol{\mu}}_{\theta}\right)\boldsymbol{\Sigma}_{x}\right\} \right)\nonumber \\
 & -\left\langle \gamma\right\rangle \mathfrak{Re}\Bigl\{\textrm{Tr}\Bigl(\textrm{diag}\left(\boldsymbol{\theta}^{j}-\hat{\boldsymbol{\mu}}_{\theta^{j}}\right)\nonumber \\
 & \hspace{1cm}\mathbf{A^{\mathit{j}}}^{H}\mathbf{A^{\mathit{j}}}\textrm{diag}\left(\boldsymbol{\theta}^{j}-\hat{\boldsymbol{\mu}}_{\theta^{j}}\right)\boldsymbol{\Sigma}_{x}\Bigr)\Bigr\}\nonumber \\
= & -\left\langle \gamma\right\rangle 2\mathfrak{Re}\left\{ \left(\boldsymbol{\theta}^{j}-\hat{\boldsymbol{\mu}}_{\theta^{j}}\right)^{T}\textrm{diag}\left(\mathbf{A^{\mathit{j}}}^{H}\mathbf{F}\left(\hat{\boldsymbol{\mu}}_{\theta}\right)\boldsymbol{\Sigma}_{x}\right)\right\} \nonumber \\
 & -\left\langle \gamma\right\rangle \mathfrak{Re}\Bigl\{\left(\boldsymbol{\theta}^{j}-\hat{\boldsymbol{\mu}}_{\theta^{j}}\right)^{T}\nonumber \\
 & \hspace{1cm}\textrm{diag}\left(\mathbf{A^{\mathit{j}}}^{H}\mathbf{A^{\mathit{j}}}\textrm{diag}\left(\boldsymbol{\theta}^{j}-\hat{\boldsymbol{\mu}}_{\theta^{j}}\right)\boldsymbol{\Sigma}_{x}\right)\Bigr\}\nonumber \\
= & -\left\langle \gamma\right\rangle 2\mathfrak{Re}\left\{ \left(\boldsymbol{\theta}^{j}-\hat{\boldsymbol{\mu}}_{\theta^{j}}\right)^{T}\textrm{diag}\left(\mathbf{A^{\mathit{j}}}^{H}\mathbf{F}\left(\hat{\boldsymbol{\mu}}_{\theta}\right)\boldsymbol{\Sigma}_{x}\right)\right\} \nonumber \\
 & -\left\langle \gamma\right\rangle \left(\boldsymbol{\theta}^{j}-\hat{\boldsymbol{\mu}}_{\theta^{j}}\right)^{T}\mathfrak{Re}\left\{ \boldsymbol{\Sigma}_{x}^{T}\odot\mathbf{A^{\mathit{j}}}^{H}\mathbf{A^{\mathit{j}}}\right\} \left(\boldsymbol{\theta}^{j}-\hat{\boldsymbol{\mu}}_{\theta^{j}}\right).\label{eq:third_term}
\end{align}
Substituting (\ref{eq:first_term}), (\ref{eq:second_term}), and
(\ref{eq:third_term}) into (\ref{eq:ln(qthetaj_middle)}), we have
\begin{align}
\ln q\left(\boldsymbol{\theta}^{j}\right)\propto & -\frac{1}{2}\left(\boldsymbol{\theta}^{j}-\hat{\boldsymbol{\mu}}_{\theta^{j}}\right)^{T}\left\langle \gamma\right\rangle \mathbf{H}_{\theta^{j}}\left(\boldsymbol{\theta}^{j}-\hat{\boldsymbol{\mu}}_{\theta^{j}}\right)\nonumber \\
 & +\left(\boldsymbol{\theta}^{j}-\hat{\boldsymbol{\mu}}_{\theta^{j}}\right)^{T}\left\langle \gamma\right\rangle \boldsymbol{g}_{\theta^{j}}\nonumber \\
 & +\ln\mathcal{N}\left(\boldsymbol{\theta}^{j};\overline{\boldsymbol{\theta}}^{j},1/\kappa^{j}\mathbf{I}_{N}\right)+\textrm{const}\nonumber \\
\propto & \ln\mathcal{N}\left(\boldsymbol{\theta}^{j};\mathbf{H}_{\theta^{j}}^{-1}\boldsymbol{g}_{\theta^{j}}+\hat{\boldsymbol{\mu}}_{\theta^{j}},\left\langle \gamma\right\rangle ^{-1}\mathbf{H}_{\theta^{j}}^{-1}\right)\nonumber \\
 & +\ln\mathcal{N}\left(\boldsymbol{\theta}^{j};\overline{\boldsymbol{\theta}}^{j},1/\kappa^{j}\mathbf{I}_{N}\right)+\textrm{const}\nonumber \\
\propto & \ln\mathcal{N}\left(\boldsymbol{\theta}^{j};\boldsymbol{\mu}_{\theta^{j}},\boldsymbol{\Sigma}_{\theta^{j}}\right),\forall j,\label{eq:lnq(thetaj)_final_apendix}
\end{align}
where the immediate variables has been defined to simplify notations:
\begin{equation}
\begin{aligned}\mathbf{H}_{\theta^{j}}= & 2\mathfrak{Re}\left\{ \left(\boldsymbol{\mu}_{x}\boldsymbol{\mu}_{x}^{H}+\boldsymbol{\Sigma}_{x}\right)^{T}\odot\left(\mathbf{A^{\mathit{j}}}^{H}\mathbf{A^{\mathit{j}}}\right)\right\} ,\forall j,\\
\boldsymbol{g}_{\theta^{j}}= & 2\mathfrak{Re}\left\{ \textrm{diag}\left(\boldsymbol{\mu}_{x}\right)^{H}\mathbf{A^{\mathit{j}}}^{H}\left(\boldsymbol{y}-\mathbf{F}\left(\hat{\boldsymbol{\mu}}_{\theta}\right)\boldsymbol{\mu}_{x}\right)\right\} \\
 & -2\mathfrak{Re}\left\{ \textrm{diag}\left(\mathbf{A^{\mathit{j}}}^{H}\mathbf{F}\left(\hat{\boldsymbol{\mu}}_{\theta}\right)\boldsymbol{\Sigma}_{x}\right)\right\} ,\forall j,
\end{aligned}
\end{equation}
and the approximate posterior mean and covariance matrix of $\boldsymbol{\theta}^{j}$
are respectively given by
\begin{equation}
\begin{aligned}\boldsymbol{\mu}_{\theta^{j}} & =\boldsymbol{\Sigma}_{\theta^{j}}\left(\left\langle \gamma\right\rangle \left(\boldsymbol{g}_{\theta^{j}}+\mathbf{H}_{\theta^{j}}\hat{\boldsymbol{\mu}}_{\theta^{j}}\right)+\kappa^{j}\overline{\boldsymbol{\theta}}^{j}\right),\forall j,\\
\boldsymbol{\Sigma}_{\theta^{j}} & =\left(\left\langle \gamma\right\rangle \mathbf{H}_{\theta^{j}}+\kappa^{j}\mathbf{I}_{N}\right)^{-1},\forall j.
\end{aligned}
\label{eq:theta_post_appendix}
\end{equation}
}

\bibliographystyle{IEEEtran}
\bibliography{others,CS_methods}

\end{document}